%% file: main.tex
\documentclass{sig-alternate}

\usepackage{mathptmx} 

\usepackage{fancyhdr}
\usepackage[normalem]{ulem}
\usepackage[hyphens]{url}
\usepackage[sort,nocompress]{cite}
\usepackage[final]{microtype}
\usepackage[keeplastbox]{flushend}
\usepackage[bookmarks=true,breaklinks=true,letterpaper=true,colorlinks,citecolor=blue,linkcolor=blue,urlcolor=blue]{hyperref}

\usepackage{amsmath,amssymb,amsfonts}
\usepackage{algorithmic}
\usepackage{algorithm}
\usepackage{graphicx}
\usepackage{textcomp}
\usepackage{xcolor}
\usepackage{balance}
\usepackage[]{mdframed}
\usepackage{tablefootnote}
\usepackage{multirow}
\usepackage{pifont}
\usepackage{soul}
\usepackage{authblk}
\usepackage{tikz}

\newcommand{\ignore}[1]{}

\usepackage{breakurl}

\usepackage{xspace}

\newcommand{\acceleratorname}{GME\xspace}
\newcommand{\cnoc}{(\textbf{cNoC})\xspace}

\newcommand{\themod}{(\textbf{MOD})\xspace}
\newcommand{\wmac}{(\textbf{WMAC})\xspace}

\newcommand{\cnocnop}{\textbf{cNoC}\xspace}
\newcommand{\labsnop}{\textbf{LABS}\xspace}
\newcommand{\themodnop}{\textbf{MOD}\xspace}
\newcommand{\wmacnop}{\textbf{WMAC}\xspace}

\usepackage{stmaryrd}
\def\BibTeX{{\rm B\kern-.05em{\sc i\kern-.025em b}\kern-.08em
    T\kern-.1667em\lower.7ex\hbox{E}\kern-.125emX}}

\usepackage{soul}
\usepackage{xcolor}

\usepackage[firstpage=true,color=black]{background}

\SetBgContents{Accepted Manuscript: IEEE/ACM MICRO 2023, Toronto, Canada}
\SetBgPosition{current page.east}
\SetBgVshift{-13.7cm}
\SetBgHshift{8.2cm}
\SetBgOpacity{0.5}
\SetBgAngle{90.0}
\SetBgScale{1.5}

\newcommand{\Z}{\mathbb{Z}}
\newcommand{\dbrac}[1]{\llbracket {#1} \rrbracket}
\newcommand{\dbm}{\llbracket \mathbf{m} \rrbracket}
\newcommand{\dbmd}{\llbracket \mathbf{m'} \rrbracket}
\newcommand{\amp}{\mathtt{A}_{\mathbf{m}}}
\newcommand{\ampd}{\mathtt{A}_{\mathbf{m'}}}
\newcommand{\bmp}{\mathtt{B}_{\mathbf{m}}}
\newcommand{\bmpd}{\mathtt{B}_{\mathbf{m'}}}

\newcommand{\pmpd}{\mathtt{P}_{\mathbf{m'}}}
\newcommand{\evkrotr}{\mathbf{evk_{\mathrm{rot}}^{(r)}}}

\newcommand{\redmemval}{$38\%$\xspace}

\title{GME: GPU-based Microarchitectural Extensions to Accelerate Homomorphic Encryption \vspace{-0.80in}}

\input{util/authors}

\begin{document}
\maketitle



\input{sections/00_abstract}
\input{sections/01_intro}
\input{sections/02_background}

\input{sections/03_arch}

\input{sections/04_exp_setup}

\input{sections/05_general}

\input{sections/06_related_work}
\input{sections/07_conclusion}

\input{sections/acknowledgement}


\bibliographystyle{IEEEtranS}
\bibliography{bibliography/biblio}

\end{document}

%% file: util/authors.tex
\numberofauthors{1}
\author{
    \alignauthor 
    Kaustubh~Shivdikar$^1$
	Yuhui~Bao$^1$
	Rashmi~Agrawal$^2$
	Michael~Shen$^1$
	Gilbert~Jonatan$^3$
	Evelio~Mora$^4$
        Alexander~Ingare$^1$
        Neal~Livesay$^1$
        Jos\'e~L.~Abell\'an$^5$
        John~Kim$^3$
        Ajay~Joshi$^2$
        David~Kaeli$^1$

        \small $^1$Northeastern University,
	\small $^2$Boston University,
 	\small $^3$KAIST,
  	\small $^4$UCAM
	\small $^5$Universidad de Murcia,

	\small \{shivdikar.k, bao.yu, shen.mich, ingare.a, n.livesay d.kaeli\}@northeastern.edu, \{rashmi23, joshi\}@bu.edu, eamora@ucam.edu, jlabellan@um.es, \{gilbertjonatan, jjk12\}@kaist.ac.kr
\vspace{-0.13in}
}


%% file: sections/00_abstract.tex

\begin{abstract}
Fully Homomorphic Encryption (FHE) enables the processing of encrypted data without decrypting it.
FHE has garnered significant attention over the past decade as it supports secure outsourcing of data processing to remote cloud services.
Despite its promise of strong data privacy and security guarantees, FHE introduces a slowdown of up to five orders of magnitude as compared to the same computation using plaintext data.
This overhead is presently a major barrier to the commercial adoption of FHE.

In this work, we leverage GPUs to accelerate FHE, capitalizing on a well-established GPU ecosystem available in the cloud. 
We propose GME, which combines three key microarchitectural extensions along with a compile-time optimization to the current AMD CDNA GPU architecture.
First, GME integrates a lightweight on-chip compute unit (CU)-side hierarchical interconnect to retain ciphertext in cache across FHE kernels, thus eliminating redundant memory transactions. 
Second, to tackle compute bottlenecks, GME introduces special \text{MOD-units} that provide native custom hardware support for modular reduction operations, one of the most commonly executed sets of operations in FHE. 
Third, by integrating the MOD-unit with our novel pipelined $64$-bit integer arithmetic cores (WMAC-units), GME further accelerates FHE workloads by $19\%$.  
Finally, we propose a Locality-Aware Block Scheduler (LABS) that exploits the temporal locality available in FHE primitive blocks.
Incorporating these microarchitectural features and compiler optimizations, we create a synergistic approach achieving average speedups of $796\times$, $14.2\times$, and $2.3\times$ over Intel Xeon CPU, NVIDIA V100 GPU, and Xilinx FPGA implementations, respectively.
\end{abstract}


%% file: sections/01_intro.tex

\section{Introduction}
\label{sec:intro}
\begin{figure}[bhtp]
	\centering
	\includegraphics[width=0.49\textwidth]{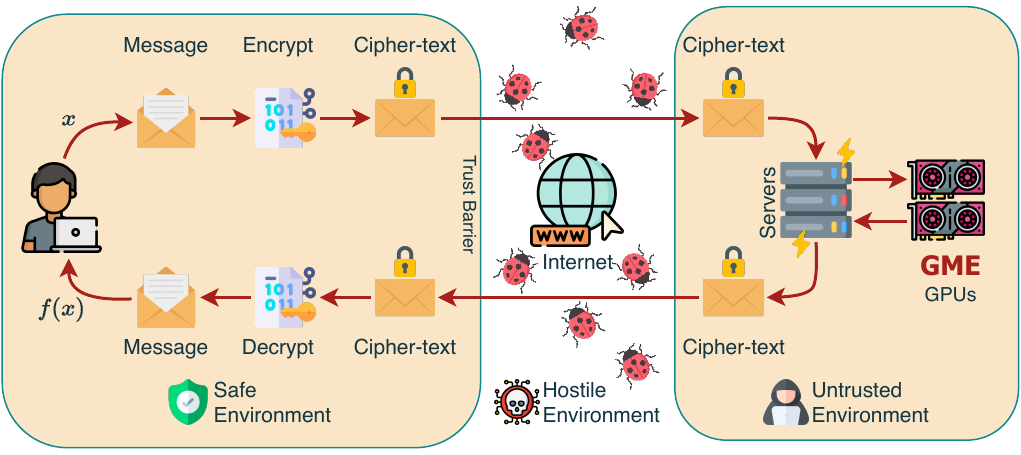}
	\caption{FHE offers a safeguard against online eavesdroppers as well as untrusted cloud services by allowing direct computation on encrypted data.}
	\label{fig:overview}
\end{figure}
Large-scale machine learning (ML) models, such as OpenAI's GPT series and DALL-E, Google AI's BERT and T5, and Facebook's RoBERTA, have made significant advances in recent years.
Unfortunately, providing public access for inference on these large-scale models leaves them susceptible to zero-day exploits~\cite{sarhan2023zero, jayaweera2021jaxed}.
These exploits expose the user data as well as the ML models to hackers for potential reverse engineering~\cite{jayaweera2021jaxed}, a concerning prospect as these models are highly valued assets for their respective companies.
For example, a recent security vulnerability in the Redis client library resulted in a data breach on ChatGPT~\cite{chatgpt_outage}, which is currently regarded as one of the leading machine learning research platforms.

In the past decade, Fully Homomorphic Encryption (FHE) has emerged as the ``holy grail'' of data privacy.  Using FHE, one can perform operations on encrypted data without decrypting it first (see Figure~\ref{fig:overview}).  
FHE adopters can offload their encrypted private data to third-party cloud service providers while preserving end-to-end privacy.  Specifically, the \textit{secret key} used for encryption by users is never disclosed to the cloud providers, thus facilitating privacy-preserving ML training and inference in an untrusted cloud setting (whether self-hosted or utilizing public cloud services)~\cite{xu2021privacy, thakkar2017video, shivdikar2015automatic}.

During its early stages, homomorphic encryption was limited by the number and types of computations, rendering it viable solely for \textit{shallow circuits}~\cite{gentry2009fully}.
In these circuits, the error would propagate and increase with each addition or multiplication operation, ultimately leading to decryption errors.  
Following Gentry's groundbreaking work~\cite{gentry2009fully}, this important limitation was resolved by using \textit{bootstrapping}~\cite{cheon2020faster}, resulting in FHE computations that permit an unlimited number of operations.
Although FHE offers significant benefits in terms of privacy preservation, it faces the challenge of being extremely slow (especially the bootstrapping operation), with performance up to five orders of magnitude slower than plaintext computing~\cite{jung2021accelerating}.

Prior studies have tried to accelerate FHE kernels by developing CPU extensions~\cite{boemer2021intel, jung2021accelerating, meftah2022towards, gentry2011implementing}, GPU libraries~\cite{shivdikar2022accelerating, al2020multi, ozcan2023homomorphic, livesay2023accelerating}, FPGA implementations~\cite{agrawal2023fab, ye2022fpga, riazi2020heax}, and custom accelerators~\cite{roy2021accelerator, kim2022bts, gupta2022memfhe}.
CPU-based solutions inherently face limitations due to their limited compute throughput~\cite{bunn2019student}, while FPGA-based solutions are constrained by their limited operating frequency and resources available on the FPGA board. ASIC-based solutions provide the most acceleration~\cite{geelen2022basalisc}, but they cannot be easily adapted to future algorithmic changes and can be fairly expensive to use in practice. Additionally, as the number of diverse domain-specific custom accelerators grows rapidly, it becomes increasingly difficult to create high-quality software libraries, compilers, drivers, and simulation tools for each accelerator in a timely manner, posing a challenge in terms of time-to-market.  Therefore, while previous work has accelerated FHE workloads, they often fall short in terms of cost-effectiveness or lack the necessary infrastructure to support large-scale deployment.

Rather than developing domain-specific custom accelerators, our work focuses on enhancing the microarchitecture of GPUs that are currently deployed in the cloud and can be easily upgraded.
This leads to a practical solution as we can readily exploit the cloud ecosystem that is built around GPUs.
On the upside, GPUs offer a large number of vector processing units, so they are a good match to capitalize on the inherent parallelism associated with FHE workloads. 
However, FHE ciphertexts are large (dozens of MB), require a massive number of integer arithmetic operations, and exhibit varying stride memory access patterns.
This imposes a true challenge for existing GPU architectures since GPUs have been historically designed to excel at executing thousands of threads in parallel (e.g., batched machine-learning workloads) featuring uniform memory access patterns and rich floating-point computations.

To bridge the wide performance gap between operating on encrypted data using FHE and operating on plaintext data in GPUs, we propose several microarchitectural features to extend the latest AMD CDNA GPU architecture. 
Specifically, our efforts are focused on improving the performance of the Residue Number System (RNS) version of the CKKS FHE scheme, as it naturally supports numerous privacy-preserving applications.  
Similar to results found in earlier studies~\cite{de2021does}, our benchmarking of CKKS FHE kernels indicates they are significantly bottlenecked by the limited main memory bandwidth.
This is because current GPUs suffer from excessive redundant memory accesses when executing FHE-based workloads. Present GPUs are ill-equipped to deal with varying stride FHE memory access patterns. According to our experiments, this can lead to a very high degree of compute unit stalls and is a primary cause of the huge performance slowdown in FHE computations on GPU-based systems.

\begin{figure}[tbp]
	\centering
	\includegraphics[width=0.475\textwidth]{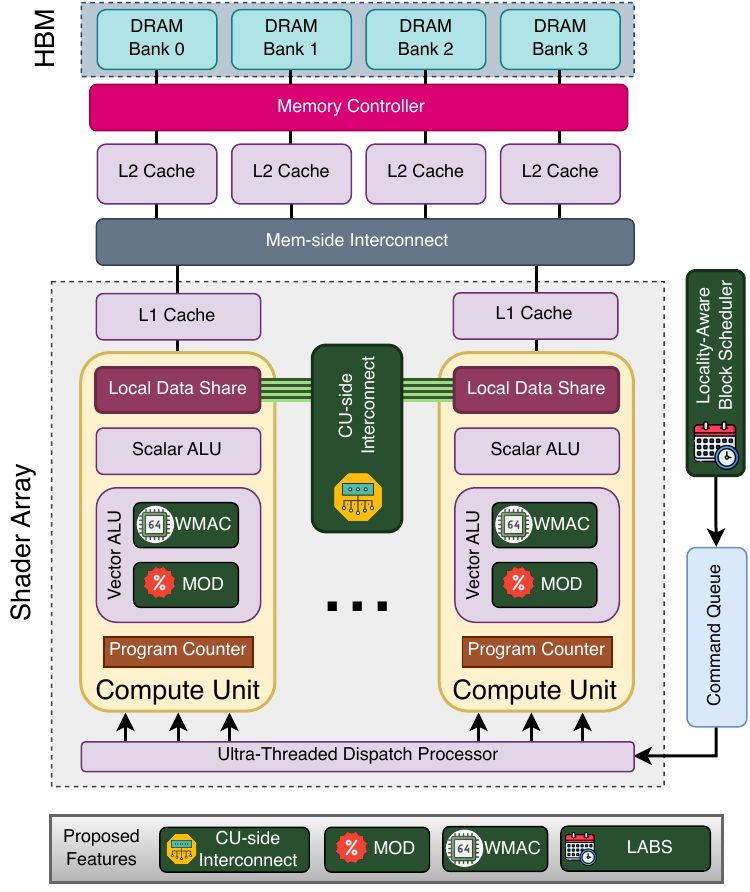}
	\caption{The four key contributions of our work (indicated in green) evaluated within the context of an AMD CDNA GPU architecture.}
	\label{fig:contributions}
\end{figure}

To address these challenges, we propose \acceleratorname, a hardware-software co-design specifically tailored to provide efficient FHE execution on the AMD CDNA GPU architecture (illustrated in Figure~\ref{fig:contributions}).  
First, we present \textit{CU-side interconnects} that allow ciphertext to be retained within the on-chip caches, thus eliminating redundant memory transactions in the FHE kernels.  Next, we optimize the most commonly executed operations present in FHE workloads (i.e., the modular reduction operations) and propose novel \textit{MOD-units}. 
To complement our \textit{MOD-units}, we introduce \textit{WMAC-units} that natively perform $64$-bit integer operations, preventing the throttling of the existing 32-bit arithmetic GPU pipelines.   Finally, in order to fully benefit from the optimizations applied to FHE kernels, we develop a Locality-Aware Block Scheduler (LABS) that enhances the temporal locality of data.  LABS is able to retain on-chip cache data across FHE blocks, utilizing block computation graphs for assistance.

To faithfully implement and evaluate \acceleratorname, we employ NaviSim~\cite{bao2022navisim}, a cycle-accurate GPU architecture simulator that accurately models the CDNA ISA~\cite{amd2020mi100isa}.
To further extend our research to capture inter-kernel optimizations, we extend the implementation of NaviSim with a block-level directed acyclic compute graph simulator called BlockSim.  
In addition, we conduct ablation studies on our microarchitectural feature implementations, enabling us to isolate each microarchitectural component and evaluate its distinct influence on the entire FHE workload.

Our contributions include:
\begin{enumerate}
\item \textit{Simulator Infrastructure:}
We introduce BlockSim, which, to the best of our knowledge, is among the first efforts to develop a simulator extension for investigating FHE microarchitecture on GPUs.
\item \textit{CU-side interconnect \cnoc:} We propose an on-chip network that interconnects on-chip memory, enabling the exploitation of the large on-chip memory capacity and support for the all-to-all communication pattern commonly found in FHE workloads.
\item \textit{GPU Microarchitecture:} We propose microarchitectural enhancements for GPUs, including ISA extensions, modular reduction operation microarchitecture, and a wide arithmetic pipeline to deliver high throughput for FHE workloads.
\item \textit{Locality-Aware Block Scheduler}: Utilizing the CU-side interconnect \cnoc, we propose a graph-based block scheduler designed to improve the temporal locality of data shared across FHE primitives.
\end{enumerate}

Our proposed improvements result in an average speedup of $14.6\times$ over the prior state-of-the-art GPU implementation~\cite{jung2021over} for HE-LR and ResNet-20 FHE workloads.
Our optimizations collectively reduce redundant computation by \redmemval, decreasing the memory pressure on DRAM.
Although the proposed optimizations can be adapted for other architectures (with minor modifications), our work primarily concentrates on AMD's CDNA microarchitecture MI100 GPU.

%% file: sections/02_background.tex
\section{Background}
\label{sec:background}
In this section, we briefly describe the AMD CDNA architecture and background of the CKKS FHE scheme.  

\subsection{AMD CDNA Architecture}
\label{subsec:cdna_arch}
\begin{figure}[tbp]
	\centering
	\includegraphics[width=0.4\textwidth]{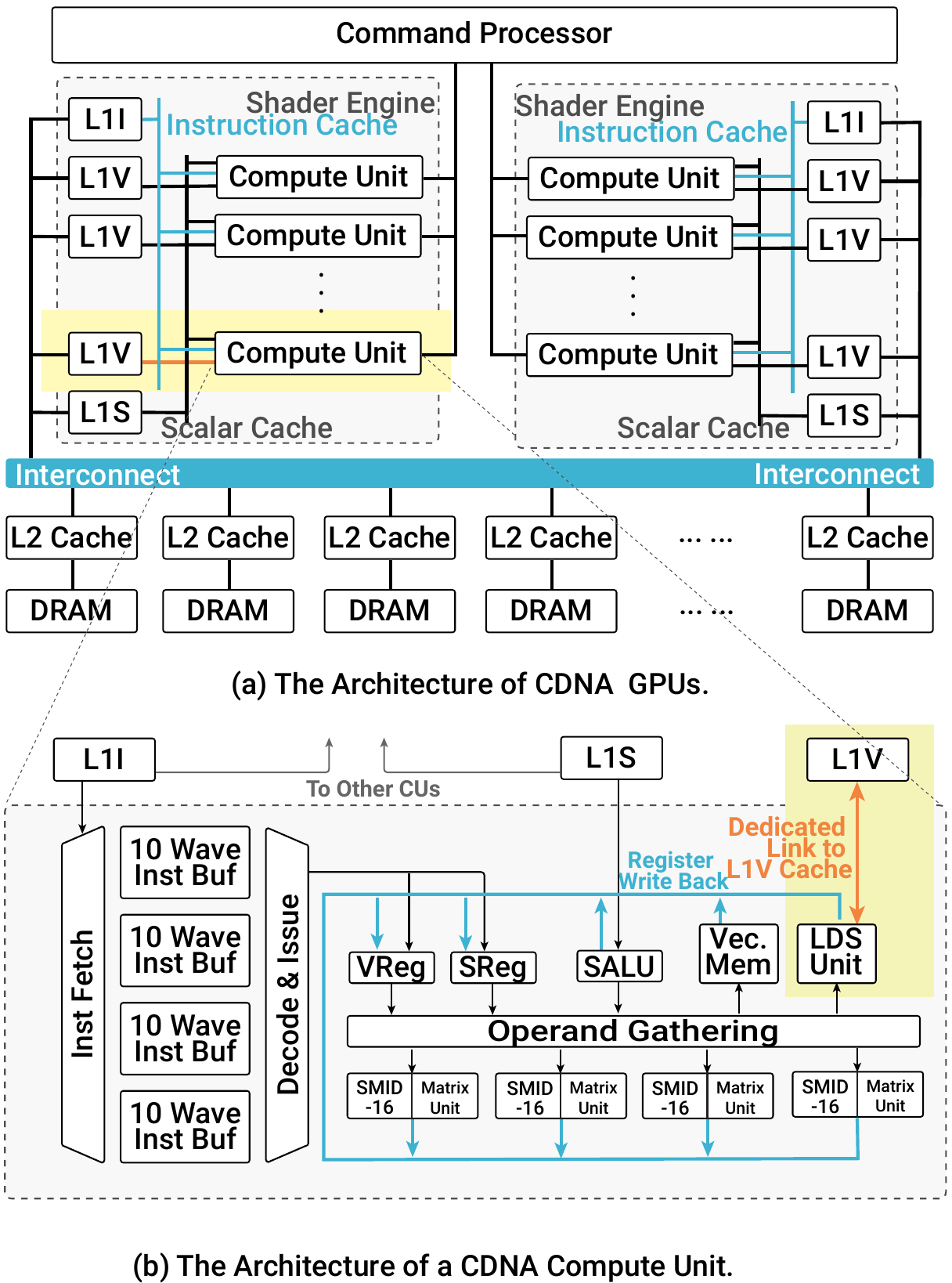}
	\caption{Architecture diagram showing the limitations of AMD GPU memory hierarchy. Each compute unit has a dedicated L1V cache and an LDS unit that cannot be shared with 
 neighboring compute units.}
	\label{fig:amd_arch}
\end{figure}

To meet the growing computation requirements of high-performance computing (HPC) and machine learning (ML) workloads, AMD introduced a new family of CDNA GPU architectures~\cite{cdna} that are used in AMD's Instinct line of accelerators. 
The CDNA architecture (see Figure~\ref{fig:amd_arch}) adopts a highly modular design that incorporates a Command Processor (CP), Shader Engines (including Compute Units and L1 caches), an interconnect connecting the core-side L1 caches to the memory-side L2 caches and DRAM.  
The CP receives requests from the driver on the CPU, including memory copying and kernel launch requests. 
The CP sends memory copying requests to the Direct Memory Access (DMA), which handles the transfer of data between the GPU and system memory. 
The CP is also responsible for breaking kernels down into work-groups and wavefronts, sending these compute tasks to Asynchronous Compute Engines (ACE), which manage the dispatch of work-groups and wavefronts on the Compute Units~(CUs).

The CDNA architecture employs the CU design from the earlier GCN architecture but enhances it with new Matrix Core Engines. 
A CU (see Figure~\ref{fig:amd_arch}) is responsible for instruction execution and data processing. 
Each CU is composed of a scheduler that can fetch and issue instructions for up to $40$ wavefronts.  
Different types of instructions are issued to different execution units, including a branch unit, scalar processing units, and vector processing units. 
The scalar processing units are responsible for executing instructions that manipulate data shared by work-items in a wavefront.  
The vector processing units include a vector memory unit, four Single-Instruction Multiple-Data (SIMD) units, and a matrix core engine.  
Each SIMD unit is equipped with $16$ single-precision Arithmetic Logic Units (ALUs), which are optimized for FP32 operations. 
The matrix core engine handles multiply-accumulate operations, supporting various datatypes (like 8-bit integers (INT8), 16-bit half-precision FP (FP16), 16-bit Brain FP (bf16), and 32-bit single-precision FP32). We cannot leverage these engines for FHE, as they work with INT8 operands that are not well-suited for FHE computations~\cite{shivdikarspeeding} (FHE workloads benefit from INT64 arithmetic pipelines). Each CU has a $64$ KB memory space called the Local Data Share (LDS), which enables low-latency communication between work-items within a work-group. LDS is analogous to shared memory in CUDA.
This memory is configured with $32$ banks to achieve low latency and high bandwidth access. 
LDS facilitates effective data sharing among work-items and acts as a software cache to minimize global memory accesses. However, a significant limitation of LDS is that CUs can only access its local LDS, and directly accessing remote LDS is not possible. 


The CDNA architecture has a two-level cache hierarchy. Each CU has a dedicated L1 vector cache. 
CUs in a Shader Engine (typically $15$ CUs) share an L1 scalar cache and an L1 instruction cache.
The second level of cache is composed of memory-side L2 caches.  Each L2 cache interfaces to a DRAM controller (typically implemented in HBM or GDDR technology). The L2 caches and the DRAM controllers are banked, allowing them to service a part of the address space. 

\subsection{CKKS FHE Scheme}
\label{subsec:ckks_scheme}
In this paper, we focus on the CKKS FHE scheme, as it can support a wide range of privacy-preserving applications by allowing operations on floating-point data.
We list the parameters that define the CKKS FHE scheme in Table~\ref{tab:ckks_params} and the corresponding values of key parameters in Table~\ref{tab:params}.
The main parameters ---i.e., $N$ and $Q$--- define the size of the ciphertext and also govern the size of the working data set that is required to be present in the on-chip memory.
The ciphertext consists of a pair of elements in the polynomial ring $R_Q = \Z_Q[x]/(x^N+1)$. 
Each element of this ring is a polynomial $\sum_{i=0}^{N-1}a_ix^i$ with ``degree-bound'' $N-1$ and coefficients $a_i$ in $\Z_Q$. 
For a message $\mathbf{m} \in \mathbb{C}^n$, we denote its encryption as $\llbracket {\mathbf{m}} \rrbracket = (\mathrm{A}_\mathbf{m}, \mathrm{B}_\mathbf{m})$ where $\mathrm{A}_\mathbf{m}$ and $\mathrm{B}_\mathbf{m}$ are the two polynomials that comprise the ciphertext.

For $128$-bit security, typical values of $N$ range from $2^{16}$ to $2^{17}$ and $\log Q$ values range from $1700$ to $2200$ bits for practical purposes.  
These large sizes of $N$ and $\log Q$ are required to maintain the security of the underlying Ring-Learning with Errors assumption~\cite{micciancio2009lattice}.
However, there are no commercially available compute systems that have hundred-bit wide or thousand-bit wide ALUs, which are necessary to process these large coefficients.
A common approach for implementing the CKKS scheme on hardware with a much smaller word length is to choose $Q$ to be a product of distinct word-sized primes $q_1,\ldots,q_{\ell}$. 
Then $\Z_Q$ can be identified with the ``product ring'' $\prod_{i=1}^l\Z_{q_i}$ via the Chinese Remainder Theorem~\cite{shoup2009}. In practice, this means that the elements of $\Z_Q$ can be represented as an $\ell$-tuple $(x_1,\ldots,x_{\ell})$ where $x_i\in\Z_{q_i}$ for each $i$. 
This representation of elements in $\Z_Q$ is referred to as the \emph{Residue Number System} (RNS) and is commonly referred to as the limbs of the ciphertext.

\begin{table}
\caption{CKKS Parameters and descriptions}
\label{tab:ckks_params}
\centering
\begin{tabular}{c l} 
    \hline
    \textbf{Param} & \textbf{Description} \\ [0.5ex] 
    \hline
    $N$ & Polynomial degree-bound\\
    $n$ & Length of the message. $n \leq \frac{N}{2}$ \\
    $Q$ & Polynomial modulus \\
    $L$ & Maximum number of limbs in a ciphertext \\
    $\mathcal{C}$ & The set $\{q_0, q_1, \dots, q_L\}$ of prime factors of $Q$\\
    $\ell$ & Number of limbs, number of factors in $Q$;\\
    $\mathsf{dnum}$ & Number of digits in the switching key \\
    $\alpha$ & Number of limbs that comprise a single digit \\ 
 & in the key-switching decomposition $\alpha = \lceil \frac{L + 1}{\mathrm{dnum}} \rceil $ \\
    $P$ & Product of extension limbs added for \\
     & raised modulus. Total extension limbs $= \alpha + 1$ \\
    $\mathsf{fftIter}$ & Multiplicative depth of bootstrapping\\
      & linear transform \\
   $\Delta$ & Scale multiplied during encryption \\
   $\mathbf{m}$ & A message vector of $n$ slots \\
   $\dbm$ & Ciphertext encrypting a message \\
   $\amp$ & A randomly sampled polynomial from message $\mathbf{m}$ \\
   $P$ & Encrypted message as a polynomial \\
   $P_m$ & Polynomial encrypting message $m$ \\
   $[P]_{q_{i}}$ & $q_i$-limb of $P$ \\
   $\mathbf{evk}$ & Evaluation key \\
   $\mathbf{evk}_{rot}^{(r)}$ & Evaluation key for \textit{HE-Rotate} block with\\
    & $(r)$ rotations \\ 
    \hline
\vspace{-6.0ex}
\end{tabular}
\end{table}

\begin{table*}
\caption{HE building blocks using CKKS}
\label{tab:ckks_blocks}
\centering
\begin{tabular*}{\textwidth}{@{\extracolsep{\fill}} l l l }
\hline \\ [-2.0ex] 
\textbf{Block} & \textbf{Computation} & \textbf{Description} \\ [0.5ex] 
\hline
\hline \\ [-1.5ex]
$\mathrm{ScalarAdd}( \dbm , c )$ &
$\dbrac{\mathbf{m + c}} = (\mathrm{B}_\mathbf{m} + \mathbf{c},\amp)$ &
Add a scalar $c$ to a ciphertext where,\\ [0.5ex]

 &  & $\mathbf{c}$ is a length-$N$ vector with every element $c$  \\ [0.5ex] 

$\mathrm{ScalarMult}( \dbm , c )$ &
$\dbrac{\mathbf{m \cdot c}} = (\mathrm{B}_\mathbf{m} \cdot \mathbf{c}, \amp \cdot \mathbf{c})$ &
Multiply a scalar by a ciphertext \\ [0.5ex] 

\hline \\ [-1.5ex]

$\mathrm{PolyAdd}( \dbm , \pmpd )$ &
$\llbracket \mathbf{m} + \mathbf{m'} \rrbracket = ( \bmp + \pmpd, \amp )$ &
Add an unencrypted polynomial \\
 &  & to a ciphertext   \\ [0.5ex] 
 
$\mathrm{PolyMult}(\dbm , \pmpd)$ &
$\llbracket \mathbf{m} \cdot \mathbf{m'} \rrbracket = (\bmp * \pmpd , \amp * \pmpd )$ &
Multiplying an unencrypted polynomial\\
 &  & with a ciphertext   \\ [0.5ex]

\hline \\ [-1.5ex]

$\mathrm{HEAdd}(\dbm , \dbmd)$ &
$\llbracket \mathbf{m} + \mathbf{m'} \rrbracket = ( \bmp + \bmpd , \amp + \ampd )$ &
Add two ciphertexts \\ [0.5ex] 

$\mathrm{HEMult}(\dbm , \dbmd , \mathbf{evk}_{\mathrm{mult}})$ &
$\llbracket \mathbf{m} \cdot \mathbf{m'} \rrbracket = \mathrm{KeySwitch}( \amp * \ampd , \mathbf{evk}_{\mathrm{mult}} ) +$ &
Multiply two ciphertexts \\ 
 & $( \bmp * \bmpd , \amp * \bmpd + \ampd * \bmp)$ & \\ [0.5ex] 

$\mathrm{HERotate}(\dbm , r , \evkrotr)$ &
$\llbracket \mathbf{m} \ll r \rrbracket = \mathrm{KeySwitch}(\psi_{r}(\amp) , \evkrotr) + $ & Circular rotate elements left by $r$ slots \\
& $( \psi_{r}(\bmp), \mathbf{0} )$ &  $\psi_{r}$ is an automorphism performed \\ [0.5ex] 

$\mathrm{HERescale}(\dbm)$ &
$\llbracket \Delta^{-1} \cdot \mathbf{m} \rrbracket = (\Delta^{-1} \bmp , \Delta^{-1}\amp)$ &
Restore the scale of a ciphertext \\
 &  & from scale $\Delta^2$ back to $\Delta$ \\ [0.5ex] 

\hline
\end{tabular*}
\end{table*}


In this work, as shown in Table~\ref{tab:params}, we choose $N=2^{16}$ and $\log Q = 1728$, meaning that our ciphertext size will be $28.3$~MB, where each polynomial in the ciphertext is ${\sim}14$~MB.
After RNS decomposition on these polynomials using a word length of $54$ bits, we get $32$ limbs in each polynomial, where each limb is $\sim 0.44$~MB large. 
The last level cache and the LDS in the AMD MI100 are $8$~MB and $7.5$~MB, respectively.  
Thus we cannot accommodate even a single ciphertext in the on-chip memory. 
At most, we can fit ${\sim}18$ limbs of a ciphertext polynomial, and as a result, we will have to perform frequent accesses to the main memory to operate on a single ciphertext. 
In addition, the large value of $N$ implies that we need to operate on $2^{16}$ coefficients for any given homomorphic operation.  
The AMD MI$100$ GPU includes $120$ CUs with $4$ SIMD units each. 
Each SIMD unit can execute $16$ threads in parallel.  
Therefore, a total of $7680$ operations (scalar additions/multiplications) can be performed in parallel.
However, we need to schedule the operations on $2^{16}$ coefficients in over eight batches ($2^{16}$ / $7680$), adding to the complexity of scheduling operations.

We list all the building blocks in the CKKS scheme in Table~\ref{tab:ckks_blocks}.  
All of the operations that form the building blocks of the CKKS scheme reduce to $64$ bit-wide scalar modular additions and scalar modular multiplications. 
The commercially available GPU architectures do not implement these wide modular arithmetic operations directly, but can emulate them via multiple arithmetic instructions, which significantly increases the amount of compute required for these operations.  
Therefore, providing native modular arithmetic units is critical to accelerating FHE computation.  
To perform modular addition over operands that are already reduced, we use the standard approach of conditional subtraction if the addition overflows the modulus. 
For generic modular multiplications, we use the modified Barrett reduction technique~\cite{shivdikar2022accelerating}. 

The $\mathrm{ScalarAdd}$ and $\mathrm{ScalarMult}$ are the two most basic building blocks that add and multiply a scalar constant to a ciphertext. 
$\mathrm{PolyAdd}$ and $\mathrm{PolyMult}$ add and multiply a plaintext polynomial to a ciphertext.
We define separate $\mathrm{ScalarAdd}$ and $\mathrm{ScalarMult}$ operations (in addition to $\mathrm{PolyAdd}$ and $\mathrm{PolyMult}$) because the scalar constant values can be fetched directly from the register file that can help save expensive main memory accesses.
Note that the $\mathrm{PolyMult}$ is followed by an $\mathrm{HERescale}$ operation to restore the scale of a ciphertext to $\Delta$ from scale $\Delta^2$.
The CKKS supports floating-point messages, so all encoded messages must include a scaling factor $\Delta$. 
This scaling factor is typically the size of one of the limbs of the ciphertext.  When multiplying messages together, this scaling factor grows as well.  The scaling factor must be shrunk down in order to avoid overflowing the ciphertext coefficient modulus.

In order to enable fast polynomial multiplication, by default, we represent polynomials as a series of $N$ evaluations at fixed roots of unity. This allows polynomial multiplication to occur in $O(N)$ time instead of $O(N^2)$ time. 
We refer to this polynomial representation as the \emph{evaluation representation}. There are certain sub-operations within the building blocks, defined in Table~\ref{tab:ckks_blocks}, that operate over 
the polynomial's \emph{coefficient representation}, which is simply a vector of its coefficients.  Moving between the two polynomial representations requires a number-theoretic transform (NTT) or inverse NTT, which is the finite field version of the fast Fourier transform (FFT).  We incorporate a merged-NTT algorithmic optimization~\cite{poppelmann}, improving spatial locality for twiddle factors as they are read sequentially.

The $\mathrm{HEAdd}$ operation is straightforward and adds the corresponding polynomials within the two ciphertexts.
However, the $\mathrm{HEMult}$ and $\mathrm{HERotate}$ operations are computationally expensive as they perform a $\mathrm{KeySwitch}$ operation after the multiplication and automorph operations, respectively.
In both the $\mathrm{HEMult}$ and $\mathrm{HERotate}$ implementations, there is an intermediate ciphertext with a decryption key that differs from the decryption key of the input ciphertexts.  
In order to change this new decryption key back to the original decryption key, we perform a key switch operation.  This operation takes in a switching key (either $\mathbf{evk}_{\mathrm{mult}}$ or $\evkrotr$) and a ciphertext $\dbm_{s}$ that is decryptable under a secret key $s$.  The output of the key switch operation is a ciphertext $\dbm_{s'}$ that encrypts the same message but is decryptable under a different key $s'$.

To incur minimal noise growth during the key switch operation, the key switch operation requires that we split the polynomial into $\mathsf{dnum}$ digits, then raise the modulus before multiplying with the switching key followed by a modulus down operation. The modulus raise and down operations operate on the coefficient representation of the polynomial, requiring us to perform expensive NTT and iNTT conversions.  Moreover, the switching keys are the same size as the ciphertext itself, requiring us to fetch ${\sim}112$~MB of data to multiply the switching keys with the ciphertext.
Thus, the key switching operation not only adds to the bulk of the compute through hundreds of NTT and iNTT operations, but also leads to memory bandwidth bottlenecks.
Finally, there exists an operation known as bootstrapping~\cite{gentry2009fully} that needs to be performed frequently to de-noise the ciphertext.  This bootstrapping operation is a sequence of the basic building blocks in the CKKS scheme, meaning that it suffers from the same compute and memory bottlenecks that exist in these building blocks, making it one of the most expensive operations.

\begin{table}
\centering
\caption{Practical parameters for our FHE operations.}
\label{tab:params}
\begin{tabular}{c c c c c c c c}
\\ [-0.5em]
\hline
\\ [-0.9em]
    $\log(q)$ & $N$ & $\log Q$ &  $L$ & $L_{boot}$  & \texttt{dnum}  & \texttt{fftIter} & $\lambda$ \\ [0.5ex] 
    \hline
    \\ [-0.9em]
     $54$ & $2^{16}$ & $1728$ & $23$ & $17$ & $3$ & $4$ & $128$ \\

     \hline
\end{tabular}
\vspace{-2.0em}
\end{table}


%% file: sections/03_arch.tex
\section{\acceleratorname Architecture}
\label{sec:uarch}
The current issue with GPUs while implementing FHE workloads is the significant disproportion in the usage of various hardware resources present on the GPUs.
As a result, specific resources such as CUs experience underutilization, while others, like HBM and on-chip caches, pose as significant bottlenecks.
In this paper, we propose to re-architect the current GPU microarchitecture and also introduce novel microarchitectural extensions that enable optimal utilization of GPU resources so as to maximize the performance of the FHE workloads running on the GPU.
We propose \textbf{\acceleratorname}, a robust set of microarchitectural features targeting AMD's CDNA architecture, unlocking the full potential of the GPU to accelerate FHE workloads over $14.2\times$ as compared to the previous comparable accelerators~\cite{jung2021over}.

In our work, we pinpoint critical bottlenecks encountered during FHE workload execution and address them progressively using four microarchitectural feature extensions.
Our on-chip CU-side hierarchical network \cnoc and the Locality Aware Block Scheduler (\textbf{LABS}) contribute to minimizing the DRAM bandwidth bottleneck. Simultaneously, our implementation of native modular reduction (\textbf{MOD}) and wider multiply-accumulate units (\textbf{WMAC}) features improve the math pipeline throughput, ensuring a streamlined data flow with evenly distributed resource utilization.
The list and impact of our contributions can be visualized in Figure~\ref{fig:contributions}.

\subsection{cNoC: CU-side interconnect}
Modern GPUs have a network-on-chip that interconnects the cores (in the case of AMD GPUs, compute units) together with the memory partitions or memory banks. On-chip communication occurs between the cores and the memory banks, not necessarily between the cores. 
In this work, we propose a new type of on-chip interconnect that we refer to as a \emph{CU-side network-on-chip \cnoc that interconnects the CUs together} -- in particular, all the CU's LDS are interconnected together with \cnoc to enable a  ``global'' LDS that can be shared between the CUs. 
By exploiting the \cnoc, the dedicated on-chip memory can be shared between cores, thus minimizing memory accesses. We also provide synchronization barriers of varying granularity to mitigate race conditions. Since the LDS is user controlled, our approach does not incur the overhead associated with cache coherence and avoids redundant cache invalidations, but comes with some extra programmer effort.
By implementing a global address space (GAS) in our GPU, we establish data sharing and form a unified GAS by combining all LDSs. The virtual address space is then mapped onto this unified GAS, with translation using a hash of the lower address bits.

\begin{figure}[tbp]
	\centering
	\includegraphics[width=0.475\textwidth]{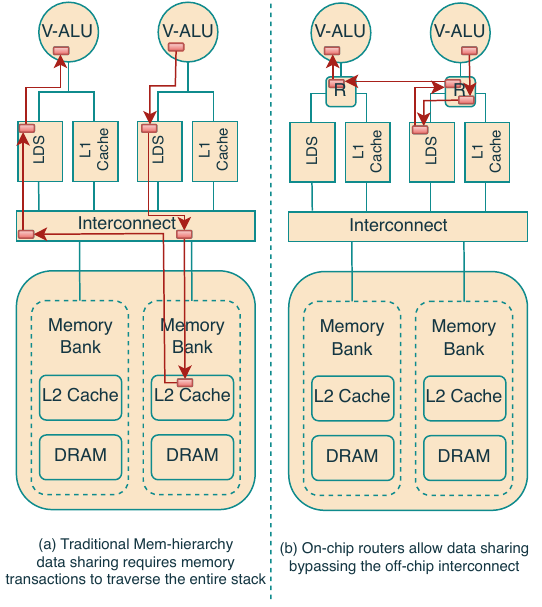}
 \vspace{-2.0em}
	\caption{Inter-CU communication: Traditional vs proposed communication with on-chip network}
	\label{fig:cu_interconnect}
\end{figure}

\begin{figure}[tbp]
	\centering
	\includegraphics[width=0.475\textwidth]{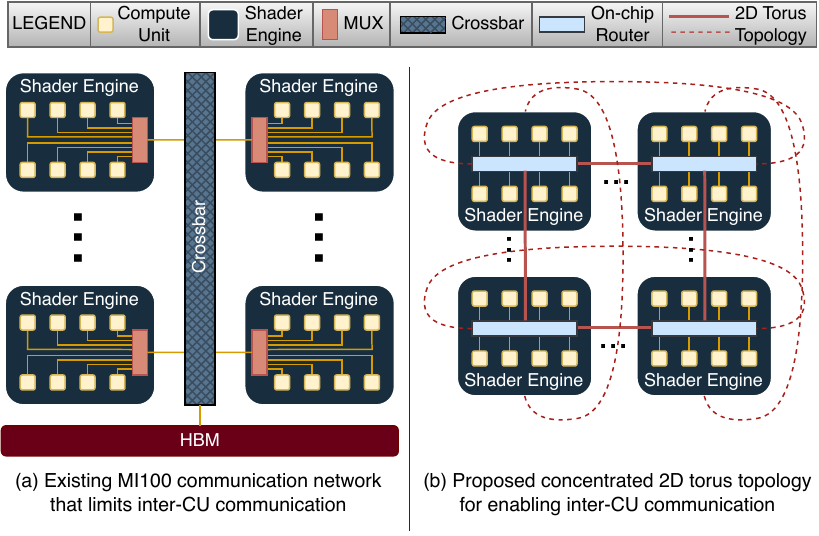}
 \vspace{-1.6em}
	\caption{Proposed hierarchical on-chip network featuring a concentrated 2D torus topology}
	\label{fig:torus}
\end{figure}

Current GPUs are designed hierarchically -- e.g., 
MI100 GPU comprises numerous compute units, with $8$ of them combined to form a \textit{Shader Engine} (seen in Figure~\ref{fig:torus}).
The proposed \cnoc takes advantage of this hierarchy, utilizing a hierarchical on-chip network (illustrated in Figure~\ref{fig:torus}) that features a single router for each \textit{Shader Engine}, connecting the eight compute units that make up a \textit{Shader Engine}.
The MI100 GPU houses $15$ \textit{Shader Engines}, resulting in a total of $120$ compute units. The routers are arranged in a $3 \times 5$ 2D grid and interconnected through a torus topology. While this \textit{concentrated-torus} topology~\cite{jerger2017chip, balfour2006design} can increase network complexity, it reduces the number of required routers (from 120 to 15), thereby minimizing the chip area needed for the network. In a concentrated-torus topology, all routers have the same degree (number of ports), creating an edge-symmetric topology that is well-suited for the all-to-all communication patterns of FHE workloads.

Figure~\ref{fig:cu_interconnect}(a) illustrates the conventional approach of data sharing, where memory transactions must traverse through the full memory hierarchy to share data between neighboring LDS. In contrast, our proposed CU-side interconnect, presented in Figure~\ref{fig:cu_interconnect}(b), incorporates on-chip routers that circumvent off-chip interconnects, improving data reuse. This results in a decrease of redundant memory operations by \redmemval, effectively supporting the all-to-all communication pattern commonly seen in FHE workloads.

\subsection{Enhancing the Vector ALU}
\textbf{Native modular reduction extension: \themod}
The existing GPU arithmetic pipeline is highly optimized for data manipulation operations like \textit{multiply}, \textit{add}, \textit{bit-shift}, and \textit{compare}. 
A wavefront executing any of these instructions takes $4$ clock cycles in a lock-step manner in the SIMD units.
In a single wavefront consisting of 64 threads, 16 threads are executed concurrently on the SIMD units during each clock cycle.
Conversely, operations like \textit{divide} and \textit{modulus} are emulated using a series of native instructions, resulting in considerably slower performance compared to their native counterparts.

As stated in Section~\ref{subsec:ckks_scheme}, the modular reduction operation, used for determining the remainder of a division, is performed after each addition and multiplication. 
As a result, optimizing modular reduction is crucial for speeding up FHE workloads.
At present, the MI100 GPU executes a modular operation through a sequence of addition, multiplication, bit shift, and conditional operations, drawing on the conventional Barrett's reduction algorithm~\cite{knezevic2010faster}. 
This operation currently takes a considerable amount of time, with the \texttt{mod-red} operation requiring an average of 46 cycles for execution on the MI100 GPU. In our study, we suggest enhancing the Vector ALU pipeline within the CDNA architecture to natively support modular reduction, which brings it down to an average of 17 cycles for each \texttt{mod-red} instruction.
We augment the CDNA instruction set architecture (ISA) with a collection of vector instructions designed to perform modular reduction operations natively after addition or multiplication operations. 
The new native modular instructions proposed include:
\begin{itemize}
    \item Native modular reduction:\\
    $\texttt{mod-red <v0,s0>}\ \ \ |\ \textbf{V}_0 = \textbf{V}_0\ \mathrm{mod}\ s_0 $
    \item Native modular addition: \\
    $\texttt{mod-add <v0,v1,s0>}\ \ \ |\ \textbf{V}_0 = (\textbf{V}_0 + \textbf{V}_1)\ \mathrm{mod}\ s_0 $ 
    \item Native modular multiplication:\\
    $\texttt{mod-mult <v0,v1,s0>}\ \ \ |\ \textbf{V}_0 = (\textbf{V}_0 \times \textbf{V}_1)\ \mathrm{mod}\ s_0 $
\end{itemize}

Modular reduction involves several comparison operations, resulting in branch divergence in GPUs. 
Our implementation is derived from an improved Barrett's reduction algorithm~\cite{shivdikar2022accelerating}.
This approach minimizes the number of comparison operations to one per modular reduction operation, significantly reducing the number of branch instructions and enhancing compute utilization.

\begin{table}
\caption{Cycle counts for 64-bit modulus instructions comparing MOD and WMAC features}
\label{tab:mod_cycle}
\centering
\begin{tabular}{l | c c c} 
    \hline
    \textbf{$\mu$-arch.} & \texttt{mod-red}  & \texttt{mod-add}  & \texttt{mod-mul} \\
    \textbf{Feature} & (cycles)$^{*}$  & (cycles)$^{*}$  & (cycles)$^{*}$ \\ [0.5ex] 
    \hline
    \textbf{Vanilla MI100}$^{\dagger}$  & 46 & 62 & 63 \\
    \textbf{MOD}$^{\Delta}$   & 26 & 18 & 38 \\
    \textbf{MOD+WMAC}          & 17 & 7 & 23 \\
    \hline
\end{tabular}

\scriptsize

\raggedright
$^{\dagger}$ Refers to the unmodified CDNA architecture of MI100 GPUs.

$^{*}$Cycle count is averaged over 10,000 modulus instructions computed on cached data (using LDS cache) and rounded to the nearest integer.

$^{\Delta}$Modular operation is computed with various compile-time prime constants as modulus incorporating compiler optimizations into the performance.

\end{table}

\textbf{Wider multiply-accumulate units \wmac:}
In the CKKS FHE scheme, we can choose to perform operations on $32$, $64$, or $128$-bit wide RNS limbs for a ciphertext. 
This limb bit width governs the operand size for the vector ALUs, impacting the number of modular addition and multiplication operations required. 
Moreover, there is an algorithmic-level performance versus precision trade-off to consider when deciding on the bit width.
If we opt for $32$-bit wide RNS limbs, we will have numerous limbs to work with, increasing the available levels~\cite{agrawal2023architecting} while simultaneously reducing the achievable precision for an application. Conversely, if we select $128$-bit RNS limbs, we will have fewer limbs to work with, resulting in a decrease in the number of available levels but result in high precision for an application. With our chosen parameters, using $128$-bit wide RNS limbs would leave us with an insufficient number of limbs to perform a single bootstrapping operation. To strike a balance between performance and precision, we choose to use $64$-bit wide RNS limbs in this work.

Most GPUs in the market natively support $16$-, $32$-, and $64$-bit floating point computations as well as $4$-, $8$-, $32$-bit integer computations. 
Unfortunately, they lack dedicated hardware support for $64$-bit integer operations, the most common operation in FHE workloads.
Instructions for processing 64-bit integer operands are emulated using multiple $32$-bit integer instructions, making them comparatively slower.
To complement our native modular reduction, which relies on $64$-bit integer operations, we add support for hardware-backed $64$-bit integer multiplier and accumulator, as well as widen the register-file size to accommodate the large ciphertexts.
Table~\ref{tab:mod_cycle} demonstrates the decrease in total cycles for each of our proposed native modular instructions in comparison to the MI100 GPU-emulated instructions in the baseline (vanilla) configuration.

Prior studies~\cite{tiwari2015performance, feng2010parallel} argued that dedicating resources to specialized 64-bit integer cores was not justifiable in terms of opportunity cost, as workloads at the time did not necessitate INT64 support, and emulation with 32-bit cores was sufficient. However, in the context of FHE, we maintain that the performance improvements attained through using an upgraded vector ALU justify the additional chip resources allocated.

\subsection{LABS: Locality-Aware Block Scheduler}

So far, our microarchitectural extensions primarily focused on optimizing individual FHE blocks. 
To better leverage these new features, we focus next on inter-block optimization opportunities, targeting the workgroup dispatcher within the CDNA architecture. 
GPU scheduling is typically managed using streams of blocks that are scheduled on compute units in a greedy manner~\cite{amdhip}. 
The presence of large GPU register files allows the scheduler to oversubscribe blocks to each compute unit. 
However, the existing scheduler within the CDNA architecture is not cognizant of inter-block data dependencies, forcing cache flushes when transitioning from one block to the next.

We propose a Locality-Aware Block Scheduler (LABS) designed to schedule blocks with shared data together, thus avoiding redundant on-chip cache flushes, specifically in the LDS. 
LABS further benefits from our set of microarchitectural enhancements, which relax the operational constraints during block scheduling and create new opportunities for optimization (for instance, the \cnoc feature enables LDS data to be globally accessible across all CUs, thereby allowing the scheduler to assign blocks to any available CU). 
To develop LABS, we employ a well-known graph-based mapping solution and frame the problem of block mapping to CUs as a compile-time Graph Partitioning Problem (GPP)~\cite{multilevel, srinivasan2016dynamic}.

\textbf{Graph Partitioning Problem: }
To develop our locality-aware block scheduler, we use two graphs. 
Let $G = G(V,E)$ represent a directed acyclic compute graph with vertices $V$ (corresponding to FHE blocks) and edges $E$ (indicating the data dependencies of the blocks). 
Similarly, let $G_a = G_a(V_a,E_a)$ denote an undirected graph with vertices $V_a$ (representing GPU compute units) and edges $E_a$ (illustrating the communication links between compute units).  
Both edge sets, $E$ and $E_a$, are assumed to be weighted, with edge weights of $E$ signifying the size of data transferred between related blocks, and $E_a$ representing the bandwidth of communication between corresponding compute units. 
We can then define $\pi : V \rightarrow V_a$ as a mapping of $V$ into $V_a$ disjoint subsets. 
Our objective is to find a mapping $\pi$ that minimizes communication overhead between compute units.

We formulate our Graph Partitioning Problem (GPP) by introducing a cost function $\Phi$. 
For a graph $G$, if it is partitioned such that $E_c$ denotes the set of edge cuts, then $\Phi$ can be expressed as the sum of the individual cut-edge weights (with $(v,w)$ representing the edge-weight of the edge connecting node $v$ to node $w$). 
The cost function $\Phi$ reflects the communication overhead associated with assigning FHE blocks to separate compute units.
The goal of the graph partitioning problem is to discover a partition that evenly distributes the load across each compute unit while minimizing the communication cost $\Phi$.
\begin{equation*}
\Phi = | E_c | = \sum\limits_{(v,w) \in E_c} | (v,w) |
\end{equation*}
In this equation, $|(v,w)|$ signifies the data transferred between FHE blocks. To partition the compute graph and prepare it for mapping onto the architecture graph, we utilize a multilevel mesh partitioning technique. For readers interested in gaining further insights into our graph partitioning implementation of the multi-level mesh partitioning algorithm, we recommend referring to the work of Walshaw and Cross~\cite{multilevel}.

\textbf{Architecture-aware mapping: }
In this work, we focus on mapping our partitioned subgraphs onto the set of compute units $V_a$, where communication costs (both latency and bandwidth) are not uniformly distributed across the network~\cite{shivdikar2021smash}.
To uniformly distribute the communication overheads across the network, we introduce a network cost function $\Gamma$.
Here, $\Gamma$ is defined as the product of individual cut-weights and their corresponding edge-weights in the architecture graph when mapped using a mapping function $\pi$. 
Formally, $\Gamma$ is described as:

\begin{equation*}
\Gamma = \sum\limits_{(v,w) \in E_c} | (v,w) | . | (\pi (v) , \pi (w)) |
\end{equation*}

In this equation, $\pi (v)$ represents the mapping of block $v$ to a compute unit from the set $V_a$, after applying the mapping function $\pi$. Additionally, $|(\pi (v) , \pi (w) )|$ represents the communication bandwidth between compute units $\pi(v)$ and $\pi(w)$. Similar to our analysis with $\Phi$, our goal is to minimize $\Gamma$. To accomplish this, we use a compile-time optimization by applying \textit{simulated annealing}, alongside mesh partitioning, to map FHE blocks onto compute units efficiently. The evaluation of performance improvements by incorporating the \labsnop is discussed further in Section~\ref{sec:eval}.

%% file: sections/04_exp_setup.tex
\section{Evaluation}
\label{sec:eval}
In this section, we first give a concise overview of the GPU simulator employed to model our microarchitectural extensions. Next, we outline the evaluation methodology assumed to assess the performance of our bootstrapping and other workload implementations. Finally, we present evaluation results.

\begin{table}
\centering
\caption{MI100 GPU Parameters}
\vspace{1.0em}
\label{tab:mi100}
\begin{tabular}{l c}
\hline
    \textbf{Parameter} & \textbf{Value} \\ [0.5ex] 
    \hline
     GPU Core Freq &  1502 MHz\\
     Process Size & 7 nm \\ 
     TFLOPS & 23.07 \\
    \hline
     Register File & 15 MB \\
     CU count & 120 \\
     L1 Vector Cache & 16 KB per CU \\
     L1 Scalar Cache & 16 KB \\
     L1 Inst Cache & 32 KB \\
     Shared L2 & 8 MB \\
     LDS & 7.5 MB \\
     GPU Memory & 32 GB HBM2 \\
     Mem Bandwidth & 1229 GB/s \\
    \hline
     Host CPU & AMD EPYC 7002 \\
     Host OS  & Ubuntu 18.04\\
     GPU Driver & AMD ROCm 5.2.5\\

     \hline
\end{tabular}
\end{table}

\subsection{The NaviSim and BlockSim Simulators}
\label{subsec:simulator}
In our work, we leverage NaviSim~\cite{bao2022navisim}, a cycle-level execution-driven GPU architecture simulator. 
NaviSim faithfully models the CDNA architecture by implementing a CDNA ISA emulator and a detailed timing simulator of all the computational components and memory hierarchy.
NaviSim utilizes the Akita simulation engine~\cite{sun2019mgpusim} to enable modularity and high-performance parallel simulation.
NaviSim is highly configurable and accurate and has been extensively validated against an AMD MI100 GPU. 
As an execution-driven simulator, NaviSim recreates the execution results of GPU instructions during simulation with the help of an instruction emulator for CDNA ISA~\cite{cdna_isa, baruah2021gnnmark}.
Currently, NaviSim supports kernels written in both OpenCL~\cite{opencl} and the HIP programming language~\cite{amdhip}. 
For our experiments, we implement our kernels using OpenCL.
NaviSim can generate a wide range of output data to facilitate performance analysis.  
For performance metrics related to individual components, NaviSim reports instruction counts, average latency spent accessing each level of cache, transaction counts for each cache, TLB transaction counts, DRAM transaction counts, and read/write data sizes. 
For low-level details, NaviSim can generate instruction traces and memory traces. 
Finally, NaviSim can produce traces using the Daisen format so that users can use Daisen, a web-based visualization tool~\cite{sun2021daisen}, to inspect the detailed behavior of each component.

We enhance NaviSim's capabilities by incorporating our new custom kernel-level simulator, BlockSim.
BlockSim is designed to enable us to identify inter-kernel optimization opportunities. With an adjustable sampling rate for performance metrics, BlockSim accelerates simulations, facilitating more efficient design space exploration.  BlockSim generates analytical models of the FHE Blocks to provide estimates for run times of various GPU configurations.
When the best design parameters are identified, NaviSim is then employed to generate cycle-accurate performance metrics.
Besides supporting FHE workloads, BlockSim serves as an essential component of NaviSim by abstracting low-level implementation details from the user, allowing them to focus on entire workloads rather than individual kernels.
BlockSim enables restructuring of the wavefront scheduler and integrates compile-time optimizations obtained from \labsnop.
We utilize AMD's CDNA architecture-based MI100 GPU to create a baseline for FHE application evaluations. 
We further validate our BlockSim findings with the MI100 GPU.

\begin{table*}
\centering
\caption{Architecture comparison of various FHE accelerators}
\label{tab:arch_config_compare}

\begin{tabular}{l | c c c c c c c c c | c c c c} 
    \multicolumn{1}{c}{\textbf{Parameters}} & 
    \multicolumn{1}{c}{\rotatebox{90}{\texttt{Lattigo}}} &
    \multicolumn{1}{c}{\rotatebox{90}{\texttt{F1 }}} &
    \multicolumn{1}{c}{\rotatebox{90}{\texttt{BTS}}} & 
    \multicolumn{1}{c}{\rotatebox{90}{\texttt{CL}}} &
    \multicolumn{1}{c}{\rotatebox{90}{\texttt{ARK}}} &
    \multicolumn{1}{c}{\rotatebox{90}{\texttt{FAB}}} &
    \multicolumn{1}{c}{\rotatebox{90}{\texttt{100x}}} &
    \multicolumn{1}{c}{\rotatebox{90}{\texttt{T-FHE}}} &
    \multicolumn{1}{c}{\rotatebox{90}{\textbf{\acceleratorname}}} &
    \multicolumn{1}{c}{\rotatebox{90}{GME-cNoC}} & 
    \multicolumn{1}{c}{\rotatebox{90}{GME-MOD}} &
    \multicolumn{1}{c}{\rotatebox{90}{GME-WMAC}} \\
    \hline
    \hline
    \\[-1.5ex]
    Technology ($nm$) & 14 & 12/14 & 7 & 12/14 & 7 & 16 & 12 & 7 & \textbf{7} \\
    Word size (bit) & 54 & 32 & 64 & 28 & 64 & 54 & 54 & 32 & \textbf{54} \\
    On-chip memory (MB) & 6 & 64 & 512 & 256 & 512 & 43 & 6 & 20.25 & \textbf{15.5} \\
    \hline

    Frequency (GHz) & 3.5 & 1.0 & 1.2 & 1.0 & 1.0 & 0.3 & 1.2 & 1.4 & \textbf{1.5} & 1.68$^\ddagger$ & 1.63$^\ddagger$ & 1.72$^\ddagger$ \\
    Area ($mm^2$) & 122 & 151.4 & 373.6 & 472.3 & 418.3 & - & 815 & 826 & \textbf{700$^*$ + 186.2}$^\dagger$ & $96.82$ & $48.27$ & $41.11$ \\
    Power ($W$) & 91 & 180.4 & 163.2 & 317 & 281.3 & 225 & 250 & 400 & \textbf{300$^*$ + 107.5}$^\dagger$ & 53.91 & 31.86 & 21.73 \\
    \hline
\end{tabular}

\vspace{0.5em}

\scriptsize

\centering
$^{*}$The CDNA architecture-based MI100 GPU chip area and power consumption are not disclosed. We display the publicly available approximated values.

$^{\dagger}$We compute the chip area and power requirements of our microarchitectural extensions using RTL components and Cadence Synthesis Solutions with the ASAP7 technology library.

$^{\ddagger}$Reported values are of maximum clock frequency $F_{max}$ that the design can sustain without violating timing constraints.

\end{table*}

\subsection{Experimental Setup}

In our experiments, we determine our baseline performance using an AMD MI100 CDNA GPU (see table~\ref{tab:mi100}).
We then iteratively introduce microarchitectural extensions and evaluate the performance benefits of each enhancement. 
We first evaluate our three microarchitectural extensions (\cnocnop, \themodnop, \wmacnop), then evaluate our compile-time optimization \labsnop, and conclude with a memory size exploration to determine the impact of on-chip memory size on FHE workloads. 
We evaluate these microarchitectural enhancements and compiler optimization using NaviSim and BlockSim.
To determine the power and area overhead of our proposed microarchitectural components, we implement them in RTL. 
Utilizing Cadence Genus Synthesis Solutions, we synthesize these RTL components targeting an ASAP7 technology library~\cite{clark2016asap7} and determine the area and power consumption for each proposed microarchitectural element.

We first evaluate our bootstrapping implementation performance, utilizing the \textit{amortized mult time per slot} metric~\cite{jung2021over}.
This metric has been used frequently in the past to perform a comparison between different bootstrapping implementations.
We can compute this metric as follows:
\begin{equation}
\mathbf{T}_{A.S.} = \frac{{\mathbf{T}_{\mathrm{boot}} + \sum_{\ell = 1}^{L - L_{\mathrm{boot}}} \mathbf{T}_{\mathrm{mult}}(\ell)} }{L - L_{\mathrm{boot}}}.\frac{1}{n}
\end{equation}
Here, $\mathbf{T}_\mathrm{boot}$ stands for total bootstrapping runtime, and $\mathrm{L}_\mathrm{boot}$ stands for the number of levels that the bootstrapping operation utilizes.
The rest of the parameters are defined in Table~\ref{tab:ckks_params}.
The parameters that we have used in our implementation have an $\mathrm{L}_\mathrm{boot} = 17$ and $n = 2^{15}$.
In addition, we analyze the performance of two workloads: HE-based logistic regression (HELR)~\cite{han2019logistic} and encrypted ResNet-$20$~\cite{lee2022low} utilizing the CIFAR-10 dataset.
For all three workloads, we evaluate the contributions of each individual FHE building block (see Table~\ref{tab:ckks_blocks}) that make up the respective workload.  
In addition, for these workloads, we report the performance benefits achieved by employing each of the proposed microarchitectural enhancements.

We also compare our implementations with other state-of-the-art CKKS accelerators, incorporating a diverse selection of CPU~\cite{bossuat2021efficient, park2023hyphen}, GPU~\cite{jung2021over, fan2023tensorfhe, park2023hyphen}, FPGA~\cite{agrawal2023fab}, and ASIC~\cite{samardzic2021f1,kim2022bts,samardzic2022craterlake,kim2022ark} platforms.\footnote{In this section, we refer to the CPU implementation as \texttt{Lattigo}, the GPU implementation as \texttt{100x}, and the CraterLake ASIC design as \texttt{CL}. For the other accelerators, we use the full names from the respective papers.}
Table~\ref{tab:arch_config_compare} presents a detailed comparison of the key architectural parameters across all the related works.
Table~\ref{tab:arch_config_compare} also showcases the distribution of chip area and power requirements for each microarchitectural enhancement of GME.
Since the maximum operating frequency $F_{max}$ of our microarchitectural enhancements ($1.63$ GHz) is greater than the typical operating frequency of the MI100 GPU ($1.5$ GHz), we do not expect our extensions to change the critical path timings of the MI100 design.
It is essential to emphasize that operating frequencies differ across various designs, a crucial factor to consider when comparing execution times in absolute terms.
Moreover, the ASIC designs make use of large on-chip memory, resulting in an expensive solution, and they are also not as flexible as CPU, GPU, and FPGA.

\vspace{1.0em}

\subsection{Results}
\label{subsec:results}

\begin{table}
\centering
\caption{Performance of various FHE building blocks}
\label{tab:building_blocks}
\begin{tabular}{l || p{0.22in} p{0.23in} p{0.27in} p{0.27in} p{0.27in}}
    \multicolumn{1}{c}{\rotatebox{90}{\textbf{}}} & 
    \multicolumn{1}{c}{\rotatebox{90}{\texttt{CMult}}} &
    \multicolumn{1}{c}{\rotatebox{90}{\texttt{HE-Add}}} &
    \multicolumn{1}{c}{\rotatebox{90}{\texttt{HE-Mult}}} &
    \multicolumn{1}{c}{\rotatebox{90}{\texttt{Rotate}}} &
    \multicolumn{1}{c}{\rotatebox{90}{\texttt{Rescale}}}  \\
    \hline
    \hline
    \\[-1.5ex]
    \texttt{HyPHEN-CPU}~\cite{park2023hyphen} ($\mu s$) & 506 & 202 & 17300 & 15500 & 3900 \\
    \texttt{100x}~\cite{jung2021over} ($\mu s$) & 130 & 160 & 2960 & 2550 & 490 \\
    
    \texttt{T-FHE}~\cite{fan2023tensorfhe} ($\mu s$) & 46 & 37 & 1131 & 1008 & 77 \\
    Baseline \texttt{MI100} ($\mu s$) & 178 & 217 & 4012 & 3473 & 681 \\
    $\mathrm{\textbf{\acceleratorname}}^*$ ($\mu s$) & $\mathbf{22}$ & $\mathbf{28}$ & $\mathbf{464}$ & $\mathbf{364}$ & $\mathbf{69}$ \\
    \hline
    \\[-1.5ex]
    Speedup over \texttt{HyPHEN} & $23\times$ & $7.2\times$ & $37.3\times$ & $42.6\times$ & $56.5\times$ \\
    Speedup over \texttt{100x} & $5.9\times$ & $5.7\times$ & $6.4\times$ & $7\times$ & $7.1\times$ \\
    Speedup over \texttt{T-FHE} & $2.1\times$ & $1.3\times$ & $2.4\times$ & $2.8\times$ & $1.1\times$ \\
    Speedup over Baseline & $8.1\times$ & $7.8\times$ & $8.6\times$ & $9.5\times$ & $9.9\times$ \\
    \hline
\end{tabular}

\vspace{0.5em}

\scriptsize

\raggedright

$^{*}$The values displayed here exclude contributions from the LABS optimization, as LABS is an \textit{inter-block} optimization, and the metrics provided are intended for individual blocks.
 \vspace{-1.0em}
\end{table}

\textbf{Performance of FHE Building Blocks:} We begin by comparing the performance of individual FHE blocks with the previous state-of-the-art GPU implementation~\cite{jung2021over}. Since these are individual FHE blocks, the reported metrics do not account for our inter-block \labsnop compiler optimization. We find that $\texttt{HEMult}$ and $\texttt{HERotate}$ are the most expensive operations, as they require key switching operations that involve the most data transfers from the main memory. The next most expensive operation is $\texttt{HERescale}$, where the runtime is dominated by the compute-intensive $\texttt{NTT}$ operations.

\begin{figure*}[tbp]
	\centering
	\includegraphics[width=1.0\textwidth]{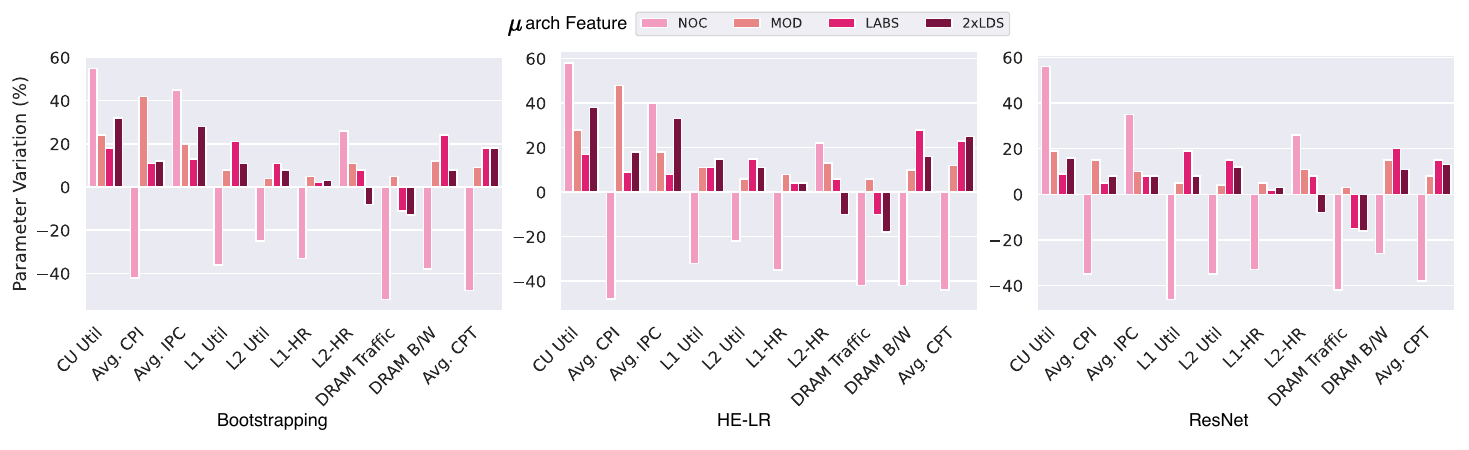}
 \vspace{-2.0em}
	\caption{Influence of individual proposed microarchitectural extension on architectural performance metrics. Metrics illustrate a cumulative profile where each enhancement builds upon the preceding set of improvements}
	\label{fig:arch_profile}
\end{figure*}

Across the five FHE blocks mentioned in Table~\ref{tab:building_blocks}, we achieve an average speedup of $6.4\times$ compared to the \texttt{100x} implementation. In particular, we see a substantial performance improvement in the most expensive operations, namely $\texttt{HEMult}$ and $\texttt{HERotate}$, as our proposed microarchitectural enhancements reduce the data transfer time by $12\times$ for both blocks. For $\texttt{HERescale}$, we manage to decrease the average memory transaction latency by $13\times$ using our microarchitectural enhancements to the on-chip network, \cnocnop. Thus making $\texttt{HERescale}$ the fastest block in comparison to \texttt{100x} GPU implementation.

\begin{figure}[tbp]
	\centering
	\includegraphics[width=0.475\textwidth]{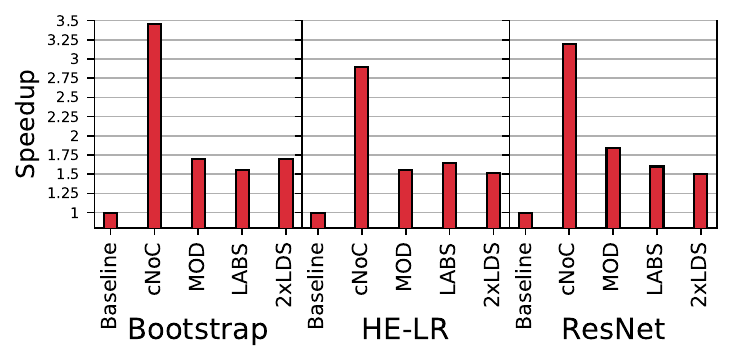}
 \vspace{-1.6em}
	\caption{Speedup achieved from each microarchitectural extension. The baseline refers to a vanilla MI100 GPU. The reported speedup is cumulative, with each microarchitectural enhancement building upon the previous ones}
	\label{fig:arch_compare}
  \vspace{-1.0em}
\end{figure}

\textbf{Impact of Microarchitectural Extensions:} Figures~\ref{fig:arch_profile} and~\ref{fig:arch_compare} highlight the impact of each of our proposed microarchitectural extensions as well as our compile-time optimizations across three different workloads, i.e., bootstrapping, HE-LR, and ResNet-$20$.

First, our proposed concentrated $2$D torus network enables ciphertexts to be preserved in on-chip memory across kernels, leading to a significant increase in compute unit utilization across workloads, thereby reducing the average cycles consumed per memory transaction (see Avg. CPT in Figure~\ref{fig:arch_profile}). 
In fact, when comparing the average number of cycles spent per memory transaction (average CPT), we observe that the ResNet-20 workload consistently displays a lower average CPT value compared to the HE-LR workload. 
This indicates a higher degree of data reuse within the ResNet-20 workload across FHE blocks as opposed to the HE-LR workload. 
With \cnocnop enhancement, as the data required from previous kernels is retained in the on-chip memory, CUs are no longer starved for data and this also results in a substantial decrease in DRAM bandwidth utilization and DRAM traffic (the total amount of data transferred from DRAM). 
The L$1$ cache utilization decreases notably across all three workloads for the \cnocnop microarchitectural enhancement. This is due to the fact that the LDS bypasses the L$1$ cache, and memory accesses to the LDS are not included in the performance metrics of the L$1$ cache.

The proposed \themodnop extension enhances the CDNA ISA by adding new instructions. 
These new instructions are complex instructions that implement commonly used operations in FHE, like \texttt{mod-red}, \texttt{mod-add}, and \texttt{mod-mult}. 
As these instructions are complex (composed of multiple sub-instructions), they consume a higher number of cycles than comparatively simpler instructions such as \texttt{mult} or \texttt{add}. 
This is the reason for the increase in the average cycles per instruction (CPI) metric shown in Figure~\ref{fig:arch_profile}. 

The compile-time \labsnop optimization in our approach further removes redundant memory transactions by scheduling blocks that share data together, thus reducing total DRAM traffic and enhancing CU utilization. 
\labsnop takes advantage of the on-chip ciphertext preservation enabled by our \cnocnop microarchitectural enhancement. 
Across bootstrapping, HE-LR, and ResNet-20 workloads, \labsnop consistently delivers an additional speedup of over $1.5\times$ on top of \cnocnop and \themodnop (See Figure~\ref{fig:arch_compare}).

\textbf{Performance Comparison:} 
We compare the performance of \acceleratorname with \texttt{100x} implementation of FHE workloads in Table~\ref{tab:prior_accel}. 
\acceleratorname surpasses the previous best GPU-based implementation for bootstrapping and HE-LR by factors of $15.7\times$ and $14.2\times$, respectively. 
Note that we do not compare the performance of ResNet-20 workload with \texttt{100x}, as they do not implement this workload.
With close to double the on-chip memory (LDS), and similar peak memory bandwidth, our microarchitectural extensions paired with our compiler optimization delivered significant performance improvement across all three FHE workloads.
\acceleratorname significantly outperforms the CPU implementation \texttt{Lattigo} by $514\times$, $1165\times$, and $427\times$ for bootstrapping, HE-LR, and ResNet-20 workloads, respectively. 
We assessed \texttt{Lattigo}'s performance by executing workloads on an Intel 8th-generation Xeon Platinum CPU with $128$~GB of DDR$4$ memory.

\begin{figure}[tb]
	\centering
	\includegraphics[width=0.475\textwidth]{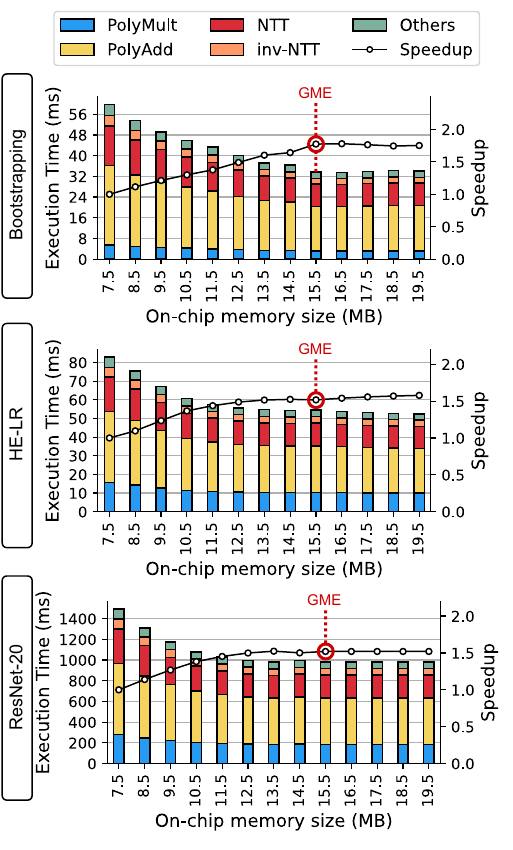}
 \vspace{-2.0em}
	\caption{Exploring the impact of on-chip memory size on FHE workload performance}
	\label{fig:spad_explore}
 \vspace{-1.0em}
\end{figure}

In addition, \acceleratorname outperforms the FPGA design implementation of FHE workloads, called \texttt{FAB}~\cite{agrawal2023fab}, by $2.7\times$ and $1.9\times$ for bootstrapping and HE-LR workloads, respectively. 
A primary factor contributing to this acceleration is the low operating frequency of FPGAs (the Alveo U$280$ used in \texttt{FAB} operates at 300MHz, while \acceleratorname cores can achieve peak frequencies of $1.5$GHz~\cite{choi2020hls}).
In their work, \texttt{FAB} scales their implementation to $8$ FPGAs for the HE-LR workload (referred to as \texttt{FAB-2}). \acceleratorname surpasses \texttt{FAB-2} by $1.4\times$. This occurs because, when the intended application cannot be accommodated on a single FPGA device, considerable communication overheads negate the 
advantages of scaling out.

However, \acceleratorname does not outperform all ASIC implementations shown in Table~\ref{tab:prior_accel}. 
While it achieves an average speedup of $18.7\times$ over \texttt{F1} for the HE-LR workload, it falls short in comparison to \texttt{BTS}, \texttt{CL}, and \texttt{ARK} due to their large on-chip memory and higher HBM bandwidths.
ASIC implementations are tailored for a single workload. Their customized designs lack flexibility, so they cannot easily accommodate multiple workloads across domains. 
Cutting-edge implementations such as \texttt{ARK}~\cite{kim2022ark} integrate the latest HBM3 technology, enabling them to utilize nearly twice the memory bandwidth available in HBM3, as compared to HBM2 used on MI100 GPUs.
CraterLake (\texttt{CL})~\cite{samardzic2022craterlake} incorporates extra physical layers (PHY) to facilitate communication between DRAM and on-chip memory, thereby enhancing the available bandwidth for FHE workloads.
In this paper, we limit our focus to an existing HBM model compatible with the CDNA architecture without modifications to the physical communication layers.

\textbf{On-chip Memory Size Exploration:} 
Finally, we look for the ideal on-chip memory (LDS) size for the FHE workload, as shown in Figure~\ref{fig:spad_explore}. 
By increasing the total LDS size from $7.5$MB (which is the current LDS size on MI100 GPU) to $15.5$MB, we achieve speedups of $1.74\times$, $1.53\times$, and $1.51\times$ for Bootstrapping, HE-LR, and ResNet-20 workloads, respectively.
However, increasing the LDS size beyond $15.5$~MB does not result in substantial speedup, as DRAM bandwidth becomes a bottleneck.

\begin{table}
\centering
\caption{HE workloads execution time comparison of proposed GME extensions with other architectures}
\vspace{1.0em}
\label{tab:prior_accel}

\begin{tabular}{l | l | p{0.27in} p{0.29in} p{0.38in} p{0.32in}} 
    \hline
    \textbf{Accelerator} & \textbf{Arch.} & \textbf{$\textbf{T}_{A.S.}$} & \textbf{Boot} & \textbf{HE-LR} & \textbf{ResNet 20} \\
     & & \textbf{($ns$)} & ($ms$) & ($ms$) & ($ms$)\\ [0.5ex] 
    \hline
    \hline
    \\ [-1.0em]
    \texttt{Lattigo}~\cite{lattigo} & CPU & $8.8e4$ & $3.9e4$ & 23293 & - \\
    \texttt{HyPHEN}~\cite{park2023hyphen} & CPU & 2110 & $2.1e4$ & -  & $3.7e4$ \\
    \hline
    \\ [-1.0em]
    \texttt{F1} ~\cite{samardzic2021f1} & ASIC & $2.6e5$ & Yes$^{\dagger}$ & 1024 & - \\
    \texttt{BTS}~\cite{kim2022bts} & ASIC & 45 & 58.9 & 28.4 & 1910 \\
    \texttt{CL}~\cite{samardzic2022craterlake} & ASIC & 17 & 4.5 & 15.2 & 321 \\
    \texttt{ARK}~\cite{kim2022ark} & ASIC & 14 & 3.7 & 7.42 & 125 \\
    \hline
    \\ [-1.0em]
    \texttt{FAB}~\cite{agrawal2023fab} & FPGA & 470 & 92.4 & 103 & - \\
    \hline
    \\ [-1.0em]
    \texttt{100x}~\cite{jung2021over} & V100 & 740 & 528 & 775  & -\\
    \texttt{HyPHEN}~\cite{park2023hyphen} & V100 & - & 830 & -  & 1400 \\
    \texttt{T-FHE}~\cite{fan2023tensorfhe} & A100 & 404 & 157 & 178  & 3793 \\
    \texttt{Baseline} & MI100 & 863 & 413 & 658  & 9989 \\
    \textbf{\acceleratorname} & \textbf{MI100+} & \textbf{74.5} & \textbf{33.63} & \textbf{54.5} & \textbf{982} \\
    \hline
\end{tabular}

\vspace{0.4em}

\scriptsize

\raggedright


$^{\dagger}$\texttt{F1} is limited to a single-slot bootstrapping, while others support packed bootstrapping.



\end{table}


\label{sec:results}

%% file: sections/05_general.tex
\section{Discussion}
In the field of accelerator design, developing general-purpose hardware is of vital importance. 
Rather than creating a custom accelerator specifically for FHE, we focus on extending the capabilities of existing GPUs to take advantage of the established ecosystems for GPUs. 
General-purpose hardware, such as GPUs, reap the benefits of versatile use of all microarchitectural elements present on the GPU.
In this section, we demonstrate the potential advantages of the proposed microarchitectural enhancements across various domains, confirming the importance of these microarchitectural features. 
Our observations are based on prior works, which highlight the potential benefits of similar optimizations across diverse workloads. 
We evaluate the influence of each optimization by examining communication overheads, high data reuse, utilizing modular reduction, or employing integer arithmetic.
Table~\ref{tab:other_workloads} presents an overview of our findings, highlighting the potential advantages of the proposed microarchitectural extensions across an array of other workloads.


The recent Hopper architecture by NVIDIA for the H100 GPU introduced a feature termed DSMEM (Distributed Shared Memory). This allows the virtual address space of shared memory to be logically spread out across various SMs (streaming multiprocessors)~\cite{elster2022nvidia}. Such a configuration promotes data sharing between SMs, similar to the \cnoc feature we introduced. However, the details of the SM-to-SM network for DSMEM are not publicly available and to the best of our knowledge, the SM-to-SM connectivity is not global but limited to the Thread Block Cluster comprised of 8 SMs. In contrast, the \cnoc proposed by us enables global connectivity to all 120 CUs in our MI100 GPU, enabling efficient all-to-all communication.
For enhancing FHE performance, it's crucial to substantially reduce the latency in SM-to-SM communication.
We aim to conduct a detailed analysis comparing the inter-SM communication overheads of the H100 GPU to those of GME in future work.

%% file: sections/06_related_work.tex
\section{Related Work}
\label{sec:related_work}

\textbf{CPU/GPU implementations:}
Several algorithmic implementations, such as \texttt{Lattigo}~\cite{mouchet2020lattigo}, SEAL~\cite{sealcrypto}, HEXL~\cite{boemer2021intel}, HEAAN~\cite{cheon2017homomorphic}, HELib~\cite{halevi2014algorithms, helib_github}, and PALISADE~\cite{polyakov_palisade}, have recently been proposed for FHE using the CKKS scheme.
Despite the efforts put forth by these libraries, a CPU-based implementation of FHE remains infeasible due to the relatively limited computational power of CPUs.

PRIFT~\cite{al2020privft} and the work by Badawi et al.~\cite{al2018high} aims to accelerate FHE using NVIDIA GPUs. 
Although they support most HE blocks, they do not accelerate bootstrapping. 
\texttt{100x}~\cite{jung2021over} speeds up all HE blocks, including bootstrapping. 
While \texttt{100x} optimizes off-chip memory transactions through \textit{kernel-fusions}, their implementation still results in 
redundant memory transactions due to partitioned on-chip memory of V100.
Locality-aware block scheduling~\cite{blockscheduling} has been proposed in GPUs to maximize locality within each core; however, LABS maximizes locality by exploiting the globally shared LDS through the proposed \cnoc.

\begin{table}
\centering
\caption{Potential benefits of proposed microarchitectural extensions across various workloads}
\vspace{1.0em}
\label{tab:other_workloads}
\begin{tabular}{ l | p{0.30in}p{0.30in}p{0.40in}p{0.32in}}
\hline
\\[-0.5ex]
    \textbf{Applications} & \textbf{NOC} & \textbf{MOD} & \textbf{WMAC} & \textbf{LABS} \\ [0.5ex] 
    \hline
    \hline \\[-1.5ex]
    AES~\cite{le2010parallel, iwai2010aes} & \ding{52} & \ding{52} & \ding{52} & \ding{52} \\
    FFT~\cite{ding2020accelerated} & \ding{52} & \ding{52} & \ding{52} & \ding{52} \\
    3D Laplace~\cite{xiao2019efficient, shin20143d} & \ding{52} & \ding{56} & \ding{52} & \ding{52} \\
    BFS~\cite{merrill2012scalable, busato2014bfs} & \ding{52} & \ding{56} & \ding{52} & \ding{117} \\
    K-Means~\cite{cuomo2019gpu} & \ding{52} & \ding{56} & \ding{56} & \ding{52} \\
    ConvNet2~\cite{li2016performance} & \ding{52} & \ding{56} & \ding{52} & \ding{117} \\
    Transformer~\cite{scao2022language, javaheripi2022litetransformersearch} & \ding{52} & \ding{56} & \ding{52} & \ding{117} \\
    Monte Carlo~\cite{lee2010debunking} & \ding{56} & \ding{56} & \ding{52} & \ding{56} \\
    N-Queens~\cite{jianli2020parallel} & \ding{56} & \ding{56} & \ding{52} & \ding{52} \\
    Black-Scholes~\cite{grauer2013accelerating} & \ding{56} & \ding{56} & \ding{52} & \ding{56} \\
    Fast Walsh~\cite{bikov2018parallel} & \ding{52} & \ding{56} & \ding{52} & \ding{52} \\
     \hline
\end{tabular}

\vspace{0.4em}

\scriptsize

\raggedright
\ding{52} Proposed optimization has the potential to significantly improve workload performance.

\ding{56} Proposed optimization is unlikely to result in notable performance improvements.

\ding{117} Further experimentation is necessary, as it is uncertain whether the proposed optimization will lead to performance improvement



\end{table}

\textbf{FPGA accelerators:}
Multiple prior efforts~\cite{riazi2020heax, roy2019fpga, kim2020hardware, kim2019fpga} have developed designs for FHE workloads. 
However, most of them either do not cover all HE primitives or only support smaller parameter sets that allow computation up to a multiplicative depth of $10$. 
HEAX~\cite{riazi2020heax} is an FPGA-based accelerator that only speeds up CKKS encrypted multiplication, with the remainder offloaded to the host processor.

FAB demonstrates performance comparable to the previous GPU implementation, \texttt{100x}~\cite{jung2021over}, and ASIC designs \texttt{BTS}~\cite{kim2022bts} and \texttt{F1}~\cite{samardzic2021f1} for certain FHE workloads. 
Although FPGAs show great potential for accelerating FHE workloads, they are limited by low operating frequencies and compute resources. 
Furthermore, 
the substantial communication overhead and the time required to program the FPGA discourages their wide-scale deployment~\cite{podobas2020survey}.

\textbf{ASIC accelerators:}
There exist several recent ASIC designs including F$1$~\cite{samardzic2021f1}, CraterLake~\cite{samardzic2022craterlake}, BTS~\cite{kim2022bts}, and ARK~\cite{kim2022ark} that accelerate the CKKS FHE scheme. 
F$1$ implementation makes use of small $N$ and $Q$ values, implementing only a single-slot bootstrapping.
BTS is the first ASIC proposal demonstrating the performance of a fully-packed CKKS bootstrapping.
CraterLake and ARK design further enhance the packed CKKS bootstrapping performance and demonstrate several orders of performance improvement across various workloads.  


%% file: sections/07_conclusion.tex
\section{Conclusion}
\label{sec:conclusion}
In this work, we present an ambitious plan for extending existing GPUs to support FHE. 
We propose three novel microarchitectural extensions followed by compiler optimization.
We suggest a 2D torus on-chip network that caters to the all-to-all communication patterns of FHE workloads.
Our native modular reduction ISA extension reduces the latency of modulus reduction operation by $43\%$.  
We enable native support for $64$-bit integer arithmetic to mitigate math pipeline throttling. 
Our proposed BlockSim simulator enhances the capabilities of the open-source GPU simulator, NaviSim, allowing for coarse-grained simulation for faster design space exploration.
Overall, comparing against previous state-of-the-art GPU implementations~\cite{jung2021over}, we obtain an average speedup of $14.6\times$ across workloads as well as outperform the CPU, the FPGA, and some ASIC implementations.

%% file: sections/acknowledgement.tex
\section*{Acknowledgments}
\vspace{-1.0em}



This research was supported in part by the Institute for Experiential AI and the NSF IUCRC Center for Hardware and Embedded Systems Security and Trust (CHEST),  NSF CNS 2312275, NSF CNS 2312276, and by Samsung Advanced Institute of Technology, Samsung Electronics Co., Ltd. Additionally, we acknowledge the financial assistance from grant RYC2021-031966-I funded by MCIN/AEI/10.13039/501100011033 and the ``European Union NextGenerationEU/PRTR.''

%% file: main.bbl
\begin{thebibliography}{10}
\providecommand{\url}[1]{#1}
\csname url@samestyle\endcsname
\providecommand{\newblock}{\relax}
\providecommand{\bibinfo}[2]{#2}
\providecommand{\BIBentrySTDinterwordspacing}{\spaceskip=0pt\relax}
\providecommand{\BIBentryALTinterwordstretchfactor}{4}
\providecommand{\BIBentryALTinterwordspacing}{\spaceskip=\fontdimen2\font plus
\BIBentryALTinterwordstretchfactor\fontdimen3\font minus \fontdimen4\font\relax}
\providecommand{\BIBforeignlanguage}[2]{{%
\expandafter\ifx\csname l@#1\endcsname\relax
\typeout{** WARNING: IEEEtranS.bst: No hyphenation pattern has been}%
\typeout{** loaded for the language `#1'. Using the pattern for}%
\typeout{** the default language instead.}%
\else
\language=\csname l@#1\endcsname
\fi
#2}}
\providecommand{\BIBdecl}{\relax}
\BIBdecl

\bibitem{agrawal2023fab}
\BIBentryALTinterwordspacing
R.~Agrawal, L.~de~Castro, G.~Yang, C.~Juvekar, R.~Yazicigil, A.~Chandrakasan, V.~Vaikuntanathan, and A.~Joshi, ``Fab: An fpga-based accelerator for bootstrappable fully homomorphic encryption,'' in \emph{2023 IEEE International Symposium on High-Performance Computer Architecture (HPCA)}.\hskip 1em plus 0.5em minus 0.4em\relax IEEE, 2023, pp. 882--895. [Online]. Available: \url{https://doi.org/10.1109/HPCA56546.2023.10070953}
\BIBentrySTDinterwordspacing

\bibitem{agrawal2023architecting}
\BIBentryALTinterwordspacing
R.~Agrawal and A.~Joshi, ``On architecting fully homomorphic encryption-based computing systems,'' 2023. [Online]. Available: \url{https://doi.org/10.1007/978-3-031-31754-5}
\BIBentrySTDinterwordspacing

\bibitem{al2020privft}
A.~Al~Badawi, L.~Hoang, C.~F. Mun, K.~Laine, and K.~M.~M. Aung, ``Privft: Private and fast text classification with homomorphic encryption,'' \emph{IEEE Access}, vol.~8, pp. 226\,544--226\,556, 2020.

\bibitem{al2020multi}
A.~Al~Badawi, B.~Veeravalli, J.~Lin, N.~Xiao, M.~Kazuaki, and A.~K.~M. Mi, ``Multi-{GPU} design and performance evaluation of homomorphic encryption on {GPU} clusters,'' \emph{IEEE Transactions on Parallel and Distributed Systems}, vol.~32, no.~2, pp. 379--391, 2020.

\bibitem{al2018high}
A.~Al~Badawi, B.~Veeravalli, C.~F. Mun, and K.~M.~M. Aung, ``High-performance fv somewhat homomorphic encryption on {GPU}s: An implementation using cuda,'' \emph{IACR Transactions on Cryptographic Hardware and Embedded Systems}, pp. 70--95, 2018.

\bibitem{amd2020mi100isa}
\BIBentryALTinterwordspacing
\emph{AMD Instinct MI100 Instruction Set Architecture}, AMD, 12 2020, reference Guide. [Online]. Available: \url{https://www.amd.com/content/dam/amd/en/documents/instinct-tech-docs/instruction-set-architectures/instinct-mi100-cdna1-shader-instruction-set-architecture.pdf}
\BIBentrySTDinterwordspacing

\bibitem{cdna_isa}
\BIBentryALTinterwordspacing
{AMD Inc.}, ``"amd instinct mi100" instruction set architecture, reference guide,'' 2020. [Online]. Available: \url{https://developer.amd.com/wp-content/resources/CDNA1_Shader_ISA_14December2020.pdf}
\BIBentrySTDinterwordspacing

\bibitem{cdna}
\BIBentryALTinterwordspacing
{AMD Inc.}, ``Introducing cdna architecture, the all-new {AMD} {GPU} architecture for the modern era of hpc \& ai,'' 2020. [Online]. Available: \url{https://www.amd.com/system/files/documents/amd-cdna-whitepaper.pdf}
\BIBentrySTDinterwordspacing

\bibitem{amdhip}
\BIBentryALTinterwordspacing
{AMD Inc.}, ``Hip programming guide,'' 2022. [Online]. Available: \url{https://rocmdocs.amd.com/en/latest/Programming_Guides/HIP-GUIDE.html}
\BIBentrySTDinterwordspacing

\bibitem{balfour2006design}
\BIBentryALTinterwordspacing
J.~Balfour and W.~J. Dally, ``Design tradeoffs for tiled cmp on-chip networks,'' in \emph{ACM International conference on supercomputing 25th anniversary volume}, 2006, pp. 390--401. [Online]. Available: \url{https://doi.org/10.1145/2591635.2667187}
\BIBentrySTDinterwordspacing

\bibitem{bao2022navisim}
\BIBentryALTinterwordspacing
Y.~Bao, Y.~Sun, Z.~Feric, M.~T. Shen, M.~Weston, J.~L. Abell{\'a}n, T.~Baruah, J.~Kim, A.~Joshi, and D.~Kaeli, ``Navisim: A highly accurate {GPU} simulator for {AMD} {RDNA} {GPU}s,'' in \emph{Proceedings of the International Conference on Parallel Architectures and Compilation Techniques}, 2022, pp. 333--345. [Online]. Available: \url{https://doi.org/10.1145/3559009.3569666}
\BIBentrySTDinterwordspacing

\bibitem{baruah2021gnnmark}
\BIBentryALTinterwordspacing
T.~Baruah, K.~Shivdikar, S.~Dong, Y.~Sun, S.~A. Mojumder, K.~Jung, J.~L. Abell{\'a}n, Y.~Ukidave, A.~Joshi, J.~Kim \emph{et~al.}, ``Gnnmark: A benchmark suite to characterize graph neural network training on gpus,'' in \emph{2021 IEEE International Symposium on Performance Analysis of Systems and Software (ISPASS)}.\hskip 1em plus 0.5em minus 0.4em\relax IEEE, 2021, pp. 13--23. [Online]. Available: \url{https://doi.org/10.1109/ISPASS51385.2021.00013}
\BIBentrySTDinterwordspacing

\bibitem{helib_github}
\BIBentryALTinterwordspacing
F.~Bergamaschi, ``Helib.'' [Online]. Available: \url{https://github.com/homenc/HElib}
\BIBentrySTDinterwordspacing

\bibitem{bikov2018parallel}
\BIBentryALTinterwordspacing
D.~Bikov and I.~Bouyukliev, ``Parallel fast walsh transform algorithm and its implementation with cuda on {GPU}s,'' \emph{Cybernetics and Information Technologies}, vol.~18, no.~5, pp. 21--43, 2018. [Online]. Available: \url{https://eprints.ugd.edu.mk/id/eprint/20026}
\BIBentrySTDinterwordspacing

\bibitem{boemer2021intel}
\BIBentryALTinterwordspacing
F.~Boemer, S.~Kim, G.~Seifu, F.~DM~de Souza, and V.~Gopal, ``Intel hexl: accelerating homomorphic encryption with intel avx512-ifma52,'' in \emph{Proceedings of the 9th on Workshop on Encrypted Computing \& Applied Homomorphic Cryptography}, 2021, pp. 57--62. [Online]. Available: \url{https://doi.org/10.1145/3474366.3486926}
\BIBentrySTDinterwordspacing

\bibitem{bossuat2021efficient}
J.-P. Bossuat, C.~Mouchet, J.~Troncoso-Pastoriza, and J.-P. Hubaux, ``Efficient bootstrapping for approximate homomorphic encryption with non-sparse keys,'' in \emph{Advances in Cryptology--EUROCRYPT 2021: 40th Annual International Conference on the Theory and Applications of Cryptographic Techniques, Zagreb, Croatia, October 17--21, 2021, Proceedings, Part I}.\hskip 1em plus 0.5em minus 0.4em\relax Springer, 2021, pp. 587--617.

\bibitem{bunn2019student}
\BIBentryALTinterwordspacing
C.~Bunn, H.~Barclay, A.~Lazarev, F.~Yusuf, J.~Fitch, J.~Booth, K.~Shivdikar, and D.~Kaeli, ``Student cluster competition 2018, team northeastern university: Reproducing performance of a multi-physics simulations of the tsunamigenic 2004 sumatra megathrust earthquake on the amd epyc 7551 architecture,'' \emph{Parallel Computing}, vol.~90, p. 102568, 2019. [Online]. Available: \url{https://doi.org/10.1016/j.parco.2019.102568}
\BIBentrySTDinterwordspacing

\bibitem{busato2014bfs}
\BIBentryALTinterwordspacing
F.~Busato and N.~Bombieri, ``Bfs-4k: an efficient implementation of bfs for kepler {GPU} architectures,'' \emph{IEEE Transactions on Parallel and Distributed Systems}, vol.~26, no.~7, pp. 1826--1838, 2014. [Online]. Available: \url{https://doi.org/10.1109/TPDS.2014.2330597}
\BIBentrySTDinterwordspacing

\bibitem{cheon2020faster}
\BIBentryALTinterwordspacing
J.~H. Cheon, K.~Han, and D.~Kim, ``Faster bootstrapping of fhe over the integers,'' in \emph{Information Security and Cryptology--ICISC 2019: 22nd International Conference, Seoul, South Korea, December 4--6, 2019, Revised Selected Papers}.\hskip 1em plus 0.5em minus 0.4em\relax Springer, 2020, pp. 242--259. [Online]. Available: \url{https://doi.org/10.1007/978-3-030-40921-0_15}
\BIBentrySTDinterwordspacing

\bibitem{cheon2017homomorphic}
J.~H. Cheon, A.~Kim, M.~Kim, and Y.~Song, ``Homomorphic encryption for arithmetic of approximate numbers,'' in \emph{Advances in Cryptology--ASIACRYPT 2017: 23rd International Conference on the Theory and Applications of Cryptology and Information Security, Hong Kong, China, December 3-7, 2017, Proceedings, Part I 23}.\hskip 1em plus 0.5em minus 0.4em\relax Springer, 2017, pp. 409--437.

\bibitem{choi2020hls}
\BIBentryALTinterwordspacing
Y.-k. Choi, Y.~Chi, J.~Wang, L.~Guo, and J.~Cong, ``When hls meets fpga hbm: Benchmarking and bandwidth optimization,'' \emph{arXiv preprint arXiv:2010.06075}, 2020. [Online]. Available: \url{https://doi.org/10.48550/arXiv.2010.06075}
\BIBentrySTDinterwordspacing

\bibitem{clark2016asap7}
\BIBentryALTinterwordspacing
L.~T. Clark, V.~Vashishtha, L.~Shifren, A.~Gujja, S.~Sinha, B.~Cline, C.~Ramamurthy, and G.~Yeric, ``Asap7: A 7-nm finfet predictive process design kit,'' \emph{Microelectronics Journal}, vol.~53, pp. 105--115, 2016. [Online]. Available: \url{https://doi.org/10.1016/j.mejo.2016.04.006}
\BIBentrySTDinterwordspacing

\bibitem{cuomo2019gpu}
\BIBentryALTinterwordspacing
S.~Cuomo, V.~De~Angelis, G.~Farina, L.~Marcellino, and G.~Toraldo, ``A {GPU}-accelerated parallel k-means algorithm,'' \emph{Computers \& Electrical Engineering}, vol.~75, pp. 262--274, 2019. [Online]. Available: \url{https://doi.org/10.1016/j.compeleceng.2017.12.002}
\BIBentrySTDinterwordspacing

\bibitem{de2021does}
\BIBentryALTinterwordspacing
L.~de~Castro, R.~Agrawal, R.~Yazicigil, A.~Chandrakasan, V.~Vaikuntanathan, C.~Juvekar, and A.~Joshi, ``Does fully homomorphic encryption need compute acceleration?'' \emph{arXiv preprint arXiv:2112.06396}, 2021. [Online]. Available: \url{https://doi.org/10.48550/arXiv.2112.06396}
\BIBentrySTDinterwordspacing

\bibitem{ding2020accelerated}
\BIBentryALTinterwordspacing
X.~Ding, Y.~Wu, Y.~Wang, J.~Z. Vilseck, and C.~L. Brooks~III, ``Accelerated cdocker with {GPU}s, parallel simulated annealing, and fast fourier transforms,'' \emph{Journal of chemical theory and computation}, vol.~16, no.~6, pp. 3910--3919, 2020. [Online]. Available: \url{https://doi.org/10.1021/acs.jctc.0c00145}
\BIBentrySTDinterwordspacing

\bibitem{elster2022nvidia}
\BIBentryALTinterwordspacing
A.~C. Elster and T.~A. Haugdahl, ``Nvidia hopper gpu and grace cpu highlights,'' \emph{Computing in Science \& Engineering}, vol.~24, no.~2, pp. 95--100, 2022. [Online]. Available: \url{https://doi.org/10.1109/MCSE.2022.3163817}
\BIBentrySTDinterwordspacing

\bibitem{fan2023tensorfhe}
\BIBentryALTinterwordspacing
S.~Fan, Z.~Wang, W.~Xu, R.~Hou, D.~Meng, and M.~Zhang, ``Tensorfhe: Achieving practical computation on encrypted data using gpgpu,'' in \emph{2023 IEEE International Symposium on High-Performance Computer Architecture (HPCA)}.\hskip 1em plus 0.5em minus 0.4em\relax IEEE, 2023, pp. 922--934. [Online]. Available: \url{https://doi.org/10.1109/HPCA56546.2023.10071017}
\BIBentrySTDinterwordspacing

\bibitem{feng2010parallel}
\BIBentryALTinterwordspacing
Z.~Feng, Z.~Zeng, and P.~Li, ``Parallel on-chip power distribution network analysis on multi-core-multi-{GPU} platforms,'' \emph{IEEE Transactions on Very Large Scale Integration (VLSI) Systems}, vol.~19, no.~10, pp. 1823--1836, 2010. [Online]. Available: \url{https://doi.org/10.1109/TVLSI.2010.2059718}
\BIBentrySTDinterwordspacing

\bibitem{geelen2022basalisc}
\BIBentryALTinterwordspacing
R.~Geelen, M.~Van~Beirendonck, H.~V. Pereira, B.~Huffman, T.~McAuley, B.~Selfridge, D.~Wagner, G.~Dimou, I.~Verbauwhede, F.~Vercauteren \emph{et~al.}, ``Basalisc: Flexible asynchronous hardware accelerator for fully homomorphic encryption,'' \emph{arXiv preprint arXiv:2205.14017}, 2022. [Online]. Available: \url{https://doi.org/10.48550/arXiv.2205.14017}
\BIBentrySTDinterwordspacing

\bibitem{gentry2009fully}
C.~Gentry, ``Fully homomorphic encryption using ideal lattices,'' in \emph{Proceedings of the forty-first annual ACM symposium on Theory of computing}, 2009, pp. 169--178.

\bibitem{gentry2011implementing}
C.~Gentry and S.~Halevi, ``Implementing gentry’s fully-homomorphic encryption scheme,'' in \emph{Advances in Cryptology--EUROCRYPT 2011: 30th Annual International Conference on the Theory and Applications of Cryptographic Techniques, Tallinn, Estonia, May 15--19, 2011. Proceedings 30}.\hskip 1em plus 0.5em minus 0.4em\relax Springer, 2011, pp. 129--148.

\bibitem{grauer2013accelerating}
\BIBentryALTinterwordspacing
S.~Grauer-Gray, W.~Killian, R.~Searles, and J.~Cavazos, ``Accelerating financial applications on the {GPU},'' in \emph{Proceedings of the 6th Workshop on General Purpose Processor Using Graphics Processing Units}, 2013, pp. 127--136. [Online]. Available: \url{https://doi.org/10.1145/2458523.2458536}
\BIBentrySTDinterwordspacing

\bibitem{gupta2022memfhe}
\BIBentryALTinterwordspacing
S.~Gupta, R.~Cammarota, and T.~{\v{S}}. Rosing, ``Memfhe: End-to-end computing with fully homomorphic encryption in memory,'' \emph{ACM Transactions on Embedded Computing Systems}, 2022. [Online]. Available: \url{https://doi.org/10.1145/3569955}
\BIBentrySTDinterwordspacing

\bibitem{halevi2014algorithms}
S.~Halevi and V.~Shoup, ``Algorithms in helib,'' in \emph{Advances in Cryptology--CRYPTO 2014: 34th Annual Cryptology Conference, Santa Barbara, CA, USA, August 17-21, 2014, Proceedings, Part I 34}.\hskip 1em plus 0.5em minus 0.4em\relax Springer, 2014, pp. 554--571.

\bibitem{han2019logistic}
K.~Han, S.~Hong, J.~H. Cheon, and D.~Park, ``Logistic regression on homomorphic encrypted data at scale,'' in \emph{Proceedings of the AAAI conference on artificial intelligence}, vol.~33, no.~01, 2019, pp. 9466--9471.

\bibitem{iwai2010aes}
\BIBentryALTinterwordspacing
K.~Iwai, T.~Kurokawa, and N.~Nisikawa, ``Aes encryption implementation on {CUDA} {GPU} and its analysis,'' in \emph{2010 First International Conference on Networking and Computing}.\hskip 1em plus 0.5em minus 0.4em\relax IEEE, 2010, pp. 209--214. [Online]. Available: \url{https://doi.org/10.1109/IC-NC.2010.49}
\BIBentrySTDinterwordspacing

\bibitem{javaheripi2022litetransformersearch}
\BIBentryALTinterwordspacing
M.~Javaheripi, G.~de~Rosa, S.~Mukherjee, S.~Shah, T.~Religa, C.~C. Teodoro~Mendes, S.~Bubeck, F.~Koushanfar, and D.~Dey, ``Litetransformersearch: Training-free neural architecture search for efficient language models,'' \emph{Advances in Neural Information Processing Systems}, vol.~35, pp. 24\,254--24\,267, 2022. [Online]. Available: \url{https://proceedings.neurips.cc/paper_files/paper/2022/hash/9949e6906be6448230cdba9a4cb2d564-Abstract-Conference.html}
\BIBentrySTDinterwordspacing

\bibitem{jayaweera2021jaxed}
\BIBentryALTinterwordspacing
M.~Jayaweera, K.~Shivdikar, Y.~Wang, and D.~Kaeli, ``Jaxed: Reverse engineering dnn architectures leveraging jit gemm libraries,'' in \emph{2021 International Symposium on Secure and Private Execution Environment Design (SEED)}.\hskip 1em plus 0.5em minus 0.4em\relax IEEE, 2021, pp. 189--202. [Online]. Available: \url{https://doi.org/10.1109/SEED51797.2021.00030}
\BIBentrySTDinterwordspacing

\bibitem{jerger2017chip}
\BIBentryALTinterwordspacing
N.~E. Jerger, T.~Krishna, and L.-S. Peh, ``On-chip networks,'' \emph{Synthesis Lectures on Computer Architecture}, vol.~12, no.~3, pp. 1--210, 2017. [Online]. Available: \url{https://picture.iczhiku.com/resource/eetop/SYieGarAzskjOvnm.pdf}
\BIBentrySTDinterwordspacing

\bibitem{jianli2020parallel}
\BIBentryALTinterwordspacing
C.~Jianli, C.~Zhikui, W.~Yuxin, and G.~He, ``Parallel genetic algorithm for n-queens problem based on message passing interface-compute unified device architecture,'' \emph{Computational Intelligence}, vol.~36, no.~4, pp. 1621--1637, 2020. [Online]. Available: \url{https://doi.org/10.1111/coin.12300}
\BIBentrySTDinterwordspacing

\bibitem{jung2021over}
W.~Jung, S.~Kim, J.~H. Ahn, J.~H. Cheon, and Y.~Lee, ``Over 100x faster bootstrapping in fully homomorphic encryption through memory-centric optimization with {GPU}s,'' \emph{IACR Transactions on Cryptographic Hardware and Embedded Systems}, pp. 114--148, 2021.

\bibitem{jung2021accelerating}
\BIBentryALTinterwordspacing
W.~Jung, E.~Lee, S.~Kim, J.~Kim, N.~Kim, K.~Lee, C.~Min, J.~H. Cheon, and J.~H. Ahn, ``Accelerating fully homomorphic encryption through architecture-centric analysis and optimization,'' \emph{IEEE Access}, vol.~9, pp. 98\,772--98\,789, 2021. [Online]. Available: \url{https://doi.org/10.1109/ACCESS.2021.3096189}
\BIBentrySTDinterwordspacing

\bibitem{opencl}
\BIBentryALTinterwordspacing
D.~R. Kaeli, P.~Mistry, D.~Schaa, and D.~P. Zhang, \emph{Heterogeneous computing with OpenCL 2.0}.\hskip 1em plus 0.5em minus 0.4em\relax Burlington,MA,USA: Morgan Kaufmann, 2015. [Online]. Available: \url{https://dahlan.unimal.ac.id/files/ebooks2/2015%203rd%20Heterogeneous%20Computing%20with%20OpenCL%202.0.pdf}
\BIBentrySTDinterwordspacing

\bibitem{kim2022ark}
J.~Kim, G.~Lee, S.~Kim, G.~Sohn, M.~Rhu, J.~Kim, and J.~H. Ahn, ``Ark: Fully homomorphic encryption accelerator with runtime data generation and inter-operation key reuse,'' in \emph{2022 55th IEEE/ACM International Symposium on Microarchitecture (MICRO)}.\hskip 1em plus 0.5em minus 0.4em\relax IEEE, 2022, pp. 1237--1254.

\bibitem{kim2022bts}
\BIBentryALTinterwordspacing
S.~Kim, J.~Kim, M.~J. Kim, W.~Jung, J.~Kim, M.~Rhu, and J.~H. Ahn, ``{BTS}: An accelerator for bootstrappable fully homomorphic encryption,'' in \emph{Proceedings of the 49th Annual International Symposium on Computer Architecture}, 2022, pp. 711--725. [Online]. Available: \url{https://doi.org/10.1145/3470496.3527415}
\BIBentrySTDinterwordspacing

\bibitem{kim2019fpga}
S.~Kim, K.~Lee, W.~Cho, J.~H. Cheon, and R.~A. Rutenbar, ``Fpga-based accelerators of fully pipelined modular multipliers for homomorphic encryption,'' in \emph{2019 International Conference on ReConFigurable Computing and FPGAs (ReConFig)}.\hskip 1em plus 0.5em minus 0.4em\relax IEEE, 2019, pp. 1--8.

\bibitem{kim2020hardware}
S.~Kim, K.~Lee, W.~Cho, Y.~Nam, J.~H. Cheon, and R.~A. Rutenbar, ``Hardware architecture of a number theoretic transform for a bootstrappable {RNS}-based homomorphic encryption scheme,'' in \emph{2020 IEEE 28th Annual International Symposium on Field-Programmable Custom Computing Machines (FCCM)}.\hskip 1em plus 0.5em minus 0.4em\relax IEEE, 2020, pp. 56--64.

\bibitem{knezevic2010faster}
\BIBentryALTinterwordspacing
M.~Knezevic, F.~Vercauteren, and I.~Verbauwhede, ``Faster interleaved modular multiplication based on barrett and montgomery reduction methods,'' \emph{IEEE Transactions on Computers}, vol.~59, no.~12, pp. 1715--1721, 2010. [Online]. Available: \url{https://doi.org/10.1109/TC.2010.93}
\BIBentrySTDinterwordspacing

\bibitem{le2010parallel}
\BIBentryALTinterwordspacing
D.~Le, J.~Chang, X.~Gou, A.~Zhang, and C.~Lu, ``Parallel aes algorithm for fast data encryption on {GPU},'' in \emph{2010 2nd international conference on computer engineering and technology}, vol.~6.\hskip 1em plus 0.5em minus 0.4em\relax IEEE, 2010, pp. V6--1. [Online]. Available: \url{https://doi.org/10.1109/ICCET.2010.5486259}
\BIBentrySTDinterwordspacing

\bibitem{lee2022low}
E.~Lee, J.-W. Lee, J.~Lee, Y.-S. Kim, Y.~Kim, J.-S. No, and W.~Choi, ``Low-complexity deep convolutional neural networks on fully homomorphic encryption using multiplexed parallel convolutions,'' in \emph{International Conference on Machine Learning}.\hskip 1em plus 0.5em minus 0.4em\relax PMLR, 2022, pp. 12\,403--12\,422.

\bibitem{blockscheduling}
M.~Lee, S.~Song, J.~Moon, J.~Kim, W.~Seo, Y.~Cho, and S.~Ryu, ``Improving gpgpu resource utilization through alternative thread block scheduling,'' in \emph{2014 IEEE 20th International Symposium on High Performance Computer Architecture (HPCA)}, 2014, pp. 260--271.

\bibitem{lee2010debunking}
\BIBentryALTinterwordspacing
V.~W. Lee, C.~Kim, J.~Chhugani, M.~Deisher, D.~Kim, A.~D. Nguyen, N.~Satish, M.~Smelyanskiy, S.~Chennupaty, P.~Hammarlund \emph{et~al.}, ``Debunking the 100x {GPU} vs. {CPU} myth: an evaluation of throughput computing on {CPU} and {GPU},'' in \emph{Proceedings of the 37th annual international symposium on Computer architecture}, 2010, pp. 451--460. [Online]. Available: \url{https://doi.org/10.1145/1815961.1816021}
\BIBentrySTDinterwordspacing

\bibitem{li2016performance}
\BIBentryALTinterwordspacing
X.~Li, G.~Zhang, H.~H. Huang, Z.~Wang, and W.~Zheng, ``Performance analysis of {GPU}-based convolutional neural networks,'' in \emph{2016 45th International conference on parallel processing (ICPP)}.\hskip 1em plus 0.5em minus 0.4em\relax IEEE, 2016, pp. 67--76. [Online]. Available: \url{https://doi.org/10.1109/ICPP.2016.15}
\BIBentrySTDinterwordspacing

\bibitem{livesay2023accelerating}
\BIBentryALTinterwordspacing
N.~Livesay, G.~Jonatan, E.~Mora, K.~Shivdikar, R.~Agrawal, A.~Joshi, J.~L. Abell{\'a}n, J.~Kim, and D.~Kaeli, ``Accelerating finite field arithmetic for homomorphic encryption on {GPU}s,'' \emph{2023 IEEE MICRO}, 2023. [Online]. Available: \url{https://doi.org/10.1109/MM.2023.3253052}
\BIBentrySTDinterwordspacing

\bibitem{meftah2022towards}
S.~Meftah, B.~H.~M. Tan, K.~M.~M. Aung, L.~Yuxiao, L.~Jie, and B.~Veeravalli, ``Towards high performance homomorphic encryption for inference tasks on {CPU}: An {MPI} approach,'' \emph{Future Generation Computer Systems}, vol. 134, pp. 13--21, 2022.

\bibitem{merrill2012scalable}
D.~Merrill, M.~Garland, and A.~Grimshaw, ``Scalable {GPU} graph traversal,'' \emph{ACM Sigplan Notices}, vol.~47, no.~8, pp. 117--128, 2012.

\bibitem{micciancio2009lattice}
D.~Micciancio and O.~Regev, ``Lattice-based cryptography,'' \emph{Post-quantum cryptography}, pp. 147--191, 2009.

\bibitem{mouchet2020lattigo}
C.~V. Mouchet, J.-P. Bossuat, J.~R. Troncoso-Pastoriza, and J.-P. Hubaux, ``Lattigo: A multiparty homomorphic encryption library in go,'' in \emph{Proceedings of the 8th Workshop on Encrypted Computing and Applied Homomorphic Cryptography}, no. CONF, 2020, pp. 64--70.

\bibitem{lattigo}
C.~V. Mouchet, J.-P. Bossuat, J.~R. Troncoso-Pastoriza, and J.-P. Hubaux, ``Lattigo v4,'' Online: \url{https://github.com/tuneinsight/lattigo}, Aug. 2022, ePFL-LDS, Tune Insight SA.

\bibitem{chatgpt_outage}
\BIBentryALTinterwordspacing
OpenAI, ``March 20 chatgpt outage: Here's what happened,'' Mar 2023. [Online]. Available: \url{https://openai.com/blog/march-20-chatgpt-outage}
\BIBentrySTDinterwordspacing

\bibitem{ozcan2023homomorphic}
A.~{\c{S}}. {\"O}zcan, C.~Ayduman, E.~R. T{\"u}rko{\u{g}}lu, and E.~Sava{\c{s}}, ``Homomorphic encryption on {GPU},'' \emph{IEEE Access}, 2023.

\bibitem{park2023hyphen}
\BIBentryALTinterwordspacing
J.~Park, D.~Kim, and J.~H. Ahn, ``Hyphen: A hybrid packing method and optimizations for homomorphic encryption-based neural network,'' 2023. [Online]. Available: \url{https://doi.org/10.48550/arXiv.2302.02407}
\BIBentrySTDinterwordspacing

\bibitem{podobas2020survey}
\BIBentryALTinterwordspacing
A.~Podobas, K.~Sano, and S.~Matsuoka, ``A survey on coarse-grained reconfigurable architectures from a performance perspective,'' \emph{IEEE Access}, vol.~8, pp. 146\,719--146\,743, 2020. [Online]. Available: \url{https://doi.org/10.1109/ACCESS.2020.3012084}
\BIBentrySTDinterwordspacing

\bibitem{polyakov_palisade}
\BIBentryALTinterwordspacing
Y.~Polyakov, ``Palisade library.'' [Online]. Available: \url{https://gitlab.com/palisade/palisade-release}
\BIBentrySTDinterwordspacing

\bibitem{poppelmann}
\BIBentryALTinterwordspacing
T.~P{\"o}ppelmann, T.~Oder, and T.~G{\"u}neysu, ``{High-Performance Ideal Lattice-Based Cryptography on 8-Bit ATxmega Microcontrollers},'' in \emph{Progress in Cryptology---LATINCRYPT}.\hskip 1em plus 0.5em minus 0.4em\relax Springer, 2015, pp. 346--365. [Online]. Available: \url{https://doi.org/10.1145/3092951}
\BIBentrySTDinterwordspacing

\bibitem{riazi2020heax}
M.~S. Riazi, K.~Laine, B.~Pelton, and W.~Dai, ``Heax: An architecture for computing on encrypted data,'' in \emph{Proceedings of the Twenty-Fifth International Conference on Architectural Support for Programming Languages and Operating Systems}, 2020, pp. 1295--1309.

\bibitem{roy2021accelerator}
S.~S. Roy, A.~C. Mert, S.~Kwon, Y.~Shin, D.~Yoo \emph{et~al.}, ``Accelerator for computing on encrypted data,'' \emph{Cryptology ePrint Archive}, 2021.

\bibitem{roy2019fpga}
S.~S. Roy, F.~Turan, K.~Jarvinen, F.~Vercauteren, and I.~Verbauwhede, ``Fpga-based high-performance parallel architecture for homomorphic computing on encrypted data,'' in \emph{2019 IEEE International symposium on high performance computer architecture (HPCA)}.\hskip 1em plus 0.5em minus 0.4em\relax IEEE, 2019, pp. 387--398.

\bibitem{samardzic2021f1}
\BIBentryALTinterwordspacing
N.~Samardzic, A.~Feldmann, A.~Krastev, S.~Devadas, R.~Dreslinski, C.~Peikert, and D.~Sanchez, ``F1: A fast and programmable accelerator for fully homomorphic encryption,'' in \emph{MICRO-54: 54th Annual IEEE/ACM International Symposium on Microarchitecture}, 2021, pp. 238--252. [Online]. Available: \url{https://doi.org/10.1145/3466752.3480070}
\BIBentrySTDinterwordspacing

\bibitem{samardzic2022craterlake}
N.~Samardzic, A.~Feldmann, A.~Krastev, N.~Manohar, N.~Genise, S.~Devadas, K.~Eldefrawy, C.~Peikert, and D.~Sanchez, ``Craterlake: a hardware accelerator for efficient unbounded computation on encrypted data,'' in \emph{Proceedings of the 49th Annual International Symposium on Computer Architecture}, 2022, pp. 173--187.

\bibitem{sarhan2023zero}
\BIBentryALTinterwordspacing
M.~Sarhan, S.~Layeghy, M.~Gallagher, and M.~Portmann, ``From zero-shot machine learning to zero-day attack detection,'' \emph{International Journal of Information Security}, pp. 1--13, 2023. [Online]. Available: \url{https://link.springer.com/article/10.1007/s10207-023-00676-0}
\BIBentrySTDinterwordspacing

\bibitem{scao2022language}
\BIBentryALTinterwordspacing
T.~L. Scao, T.~Wang, D.~Hesslow, L.~Saulnier, S.~Bekman, M.~S. Bari, S.~Bideman, H.~Elsahar, N.~Muennighoff, J.~Phang \emph{et~al.}, ``What language model to train if you have one million {GPU} hours?'' \emph{arXiv preprint arXiv:2210.15424}, 2022. [Online]. Available: \url{https://doi.org/10.48550/arXiv.2210.15424}
\BIBentrySTDinterwordspacing

\bibitem{sealcrypto}
``{M}icrosoft {SEAL} (release 4.1),'' \url{https://github.com/Microsoft/SEAL}, Jan. 2023, microsoft Research, Redmond, WA.

\bibitem{shin20143d}
\BIBentryALTinterwordspacing
J.~Shin, W.~Ha, H.~Jun, D.-J. Min, and C.~Shin, ``3d laplace-domain full waveform inversion using a single {GPU} card,'' \emph{Computers \& Geosciences}, vol.~67, pp. 1--13, 2014. [Online]. Available: \url{https://doi.org/10.1016/j.cageo.2014.02.006}
\BIBentrySTDinterwordspacing

\bibitem{shivdikar2021smash}
\BIBentryALTinterwordspacing
K.~Shivdikar, ``\BIBforeignlanguage{English}{Smash: Sparse matrix atomic scratchpad hashing},'' Ph.D. dissertation, 2021, copyright - Database copyright ProQuest LLC; ProQuest does not claim copyright in the individual underlying works; Last updated - 2023-03-07. [Online]. Available: \url{https://www.researchgate.net/publication/352018010_SMASH_Sparse_Matrix_Atomic_Scratchpad_Hashing}
\BIBentrySTDinterwordspacing

\bibitem{shivdikar2022accelerating}
\BIBentryALTinterwordspacing
K.~Shivdikar, G.~Jonatan, E.~Mora, N.~Livesay, R.~Agrawal, A.~Joshi, J.~L. Abell{\'a}n, J.~Kim, and D.~Kaeli, ``Accelerating polynomial multiplication for homomorphic encryption on {GPU}s,'' in \emph{2022 IEEE International Symposium on Secure and Private Execution Environment Design (SEED)}.\hskip 1em plus 0.5em minus 0.4em\relax IEEE, 2022, pp. 61--72. [Online]. Available: \url{https://doi.org/10.1109/SEED55351.2022.00013}
\BIBentrySTDinterwordspacing

\bibitem{shivdikar2015automatic}
\BIBentryALTinterwordspacing
K.~Shivdikar, A.~Kak, and K.~Marwah, ``Automatic image annotation using a hybrid engine,'' in \emph{2015 Annual IEEE India Conference (INDICON)}.\hskip 1em plus 0.5em minus 0.4em\relax IEEE, 2015, pp. 1--6. [Online]. Available: \url{https://doi.org/10.1109/INDICON.2015.7443338}
\BIBentrySTDinterwordspacing

\bibitem{shivdikarspeeding}
K.~Shivdikar, K.~Paneri, and D.~Kaeli, ``Speeding up dnns using hpl based fine-grained tiling for distributed multi-gpu training.''

\bibitem{shoup2009}
\BIBentryALTinterwordspacing
V.~Shoup, \emph{A computational introduction to number theory and algebra}.\hskip 1em plus 0.5em minus 0.4em\relax Cambridge University Press, 2009. [Online]. Available: \url{https://shoup.net/ntb/ntb-v2.pdf}
\BIBentrySTDinterwordspacing

\bibitem{srinivasan2016dynamic}
\BIBentryALTinterwordspacing
M.~Srinivasan, A.~Kak, K.~Shivdikar, and C.~Warty, ``Dynamic power allocation using stackelberg game in a wireless sensor network,'' in \emph{2016 IEEE Aerospace Conference}.\hskip 1em plus 0.5em minus 0.4em\relax IEEE, 2016, pp. 1--10. [Online]. Available: \url{https://doi.org/10.1109/AERO.2016.7500918}
\BIBentrySTDinterwordspacing

\bibitem{sun2019mgpusim}
\BIBentryALTinterwordspacing
Y.~Sun, T.~Baruah, S.~A. Mojumder, S.~Dong, X.~Gong, S.~Treadway, Y.~Bao, S.~Hance, C.~McCardwell, V.~Zhao, H.~Barclay, A.~K. Ziabari, Z.~Chen, R.~Ubal, J.~L. Abell\'{a}n, J.~Kim, A.~Joshi, and D.~Kaeli, ``{MGPUSim}: Enabling multi-{GPU} performance modeling and optimization,'' in \emph{Proceedings of the 46th International Symposium on Computer Architecture}, ser. ISCA '19.\hskip 1em plus 0.5em minus 0.4em\relax New York, NY, USA: Association for Computing Machinery, 2019, p. 197–209. [Online]. Available: \url{https://doi.org/10.1145/3307650.3322230}
\BIBentrySTDinterwordspacing

\bibitem{sun2021daisen}
Y.~Sun, Y.~Zhang, A.~Mosallaei, M.~D. Shah, C.~Dunne, and D.~Kaeli, ``Daisen: A framework for visualizing detailed {GPU} execution,'' \emph{Eurographics Conference on Visualization}, vol.~40, no.~3, pp. 239--250, 2021.

\bibitem{thakkar2017video}
\BIBentryALTinterwordspacing
S.~Thakkar, K.~Shivdikar, and C.~Warty, ``Video steganography using encrypted payload for satellite communication,'' in \emph{2017 IEEE Aerospace Conference}.\hskip 1em plus 0.5em minus 0.4em\relax IEEE, 2017, pp. 1--11. [Online]. Available: \url{https://doi.org/10.1109/AERO.2017.7943978}
\BIBentrySTDinterwordspacing

\bibitem{tiwari2015performance}
\BIBentryALTinterwordspacing
A.~Tiwari, K.~Keipert, A.~Jundt, J.~Peraza, S.~S. Leang, M.~Laurenzano, M.~S. Gordon, and L.~Carrington, ``Performance and energy efficiency analysis of 64-bit arm using gamess,'' in \emph{Proceedings of the 2nd International Workshop on Hardware-Software Co-Design for High Performance Computing}, 2015, pp. 1--10. [Online]. Available: \url{https://doi.org/10.1145/2834899.2834905}
\BIBentrySTDinterwordspacing

\bibitem{multilevel}
\BIBentryALTinterwordspacing
C.~Walshaw and M.~Cross, ``Multilevel mesh partitioning for heterogeneous communication networks,'' \emph{Future generation computer systems}, vol.~17, no.~5, pp. 601--623, 2001. [Online]. Available: \url{https://doi.org/10.1016/S0167-739X(00)00107-2}
\BIBentrySTDinterwordspacing

\bibitem{xiao2019efficient}
\BIBentryALTinterwordspacing
L.~Xiao, G.~Yang, K.~Zhao, and G.~Mei, ``Efficient parallel algorithms for 3d laplacian smoothing on the {GPU},'' \emph{Applied Sciences}, vol.~9, no.~24, p. 5437, 2019. [Online]. Available: \url{https://doi.org/10.3390/app9245437}
\BIBentrySTDinterwordspacing

\bibitem{xu2021privacy}
\BIBentryALTinterwordspacing
R.~Xu, N.~Baracaldo, and J.~Joshi, ``Privacy-preserving machine learning: Methods, challenges and directions,'' \emph{arXiv preprint arXiv:2108.04417}, 2021. [Online]. Available: \url{https://doi.org/10.48550/arXiv.2108.04417}
\BIBentrySTDinterwordspacing

\bibitem{ye2022fpga}
T.~Ye, R.~Kannan, and V.~K. Prasanna, ``Fpga acceleration of fully homomorphic encryption over the torus,'' in \emph{2022 IEEE High Performance Extreme Computing Conference (HPEC)}.\hskip 1em plus 0.5em minus 0.4em\relax IEEE, 2022, pp. 1--7.

\end{thebibliography}
